\documentclass[prb,twocolumn, amsmath,amssymb,superscriptaddress,longbibliography]{revtex4-1} 

\usepackage{amsfonts,amsmath,amssymb,graphicx}

\usepackage{color,times}

\definecolor{darkblue}{rgb}{0, 0, 0.8}
\usepackage[colorlinks=true, breaklinks=true, linkcolor=red, citecolor=darkblue, urlcolor=darkblue]{hyperref}

\newcommand{\ket}[1]{\left|#1\right\rangle}
\newcommand{\bra}[1]{\left\langle#1\right|}

\newcommand{\E}{{\rm e}}

\definecolor{ctl}{rgb}{0.8, 0.1, 0.6}

\newcommand{\figref}[1]{Fig.~\ref{#1}}
\renewcommand{\eqref}[1]{Eq.~(\ref{#1})}

\begin{document}	
\title{
Programmable quantum simulation of 2D antiferromagnets with hundreds of Rydberg atoms}
\author{Pascal Scholl$^*$}
\affiliation{Universit\'e Paris-Saclay, Institut d'Optique Graduate School,\\
CNRS, Laboratoire Charles Fabry, 91127 Palaiseau Cedex, France}
\author{Michael Schuler$^*$}
\affiliation{Vienna Center for Quantum Science and Technology, Atominstitut, TU Wien, 1040 Wien, Austria}
\author{Hannah~J.~Williams$^*$}
\affiliation{Universit\'e Paris-Saclay, Institut d'Optique Graduate School,\\
	CNRS, Laboratoire Charles Fabry, 91127 Palaiseau Cedex, France}
\author{Alexander~A.~Eberharter$^*$}
\affiliation{Institut f\"ur Theoretische Physik, Universit\"at Innsbruck, A-6020 Innsbruck, Austria}
\author{Daniel Barredo}
\affiliation{Universit\'e Paris-Saclay, Institut d'Optique Graduate School,\\
	CNRS, Laboratoire Charles Fabry, 91127 Palaiseau Cedex, France}
\affiliation{Nanomaterials and Nanotechnology Research Center (CINN‑CSIC), Universidad de Oviedo (UO), Principado de Asturias, 33940 El Entrego, Spain}
\author{Kai-Niklas Schymik}
\affiliation{Universit\'e Paris-Saclay, Institut d'Optique Graduate School,\\
	CNRS, Laboratoire Charles Fabry, 91127 Palaiseau Cedex, France}
\author{Vincent Lienhard}
\affiliation{Universit\'e Paris-Saclay, Institut d'Optique Graduate School,\\
	CNRS, Laboratoire Charles Fabry, 91127 Palaiseau Cedex, France}
\author{Louis-Paul Henry}
\affiliation{Zentrum für Optische Quantentechnologien and Institut für Laser-Physik, Universität Hamburg, 22761 Hamburg, Germany}
\author{Thomas~C.~Lang}
\affiliation{Institut f\"ur Theoretische Physik, Universit\"at Innsbruck, A-6020 Innsbruck, Austria}
\author{Thierry Lahaye}
\affiliation{Universit\'e Paris-Saclay, Institut d'Optique Graduate School,\\
	CNRS, Laboratoire Charles Fabry, 91127 Palaiseau Cedex, France}
\author{Andreas~M.~L\"auchli}
\affiliation{Institut f\"ur Theoretische Physik, Universit\"at Innsbruck, A-6020 Innsbruck, Austria}
\author{Antoine Browaeys}
\affiliation{Universit\'e Paris-Saclay, Institut d'Optique Graduate School,\\
	CNRS, Laboratoire Charles Fabry, 91127 Palaiseau Cedex, France}

\date{\today}
\maketitle
	
\textbf{Quantum simulation using synthetic systems is a promising route to 
solve outstanding quantum many-body problems in regimes where 
other approaches, including numerical ones, fail\cite{GeorgescuRMP}. 
Many platforms are being developed towards this goal, in particular based on trapped 
ions\cite{Blatt2012, Monroe2020,Garttner2017}, 
superconducting circuits\cite{Song2017,King2018,Morten2020}, neutral atoms\cite{Bloch08,Bloch12,Gross17,Browaeys2020} 
or molecules\cite{Yan2013,Zhou2011}. 
All of which face two key challenges: (i) scaling up the ensemble size, whilst retaining high quality control over the parameters and (ii) certifying the outputs for these large systems.
Here, we use programmable arrays of individual atoms trapped in optical tweezers, 
with interactions controlled by laser-excitation to Rydberg states\cite{Browaeys2020}
to implement an iconic many-body problem, the antiferromagnetic 2D transverse field Ising model.
We push this platform to an unprecedented regime with up 
to 196 atoms manipulated with high fidelity.  
We probe the antiferromagnetic order by dynamically 
tuning the parameters of the Hamiltonian. 
We illustrate the versatility of our platform by exploring various system sizes 
on two qualitatively different geometries, square and triangular arrays.
We obtain good agreement
with numerical calculations up to a computationally feasible size ($\mathbf{\sim 100}$ particles). 
This work demonstrates that our platform can be readily used to address open questions in many-body physics.}
\begin{figure*}
	\includegraphics{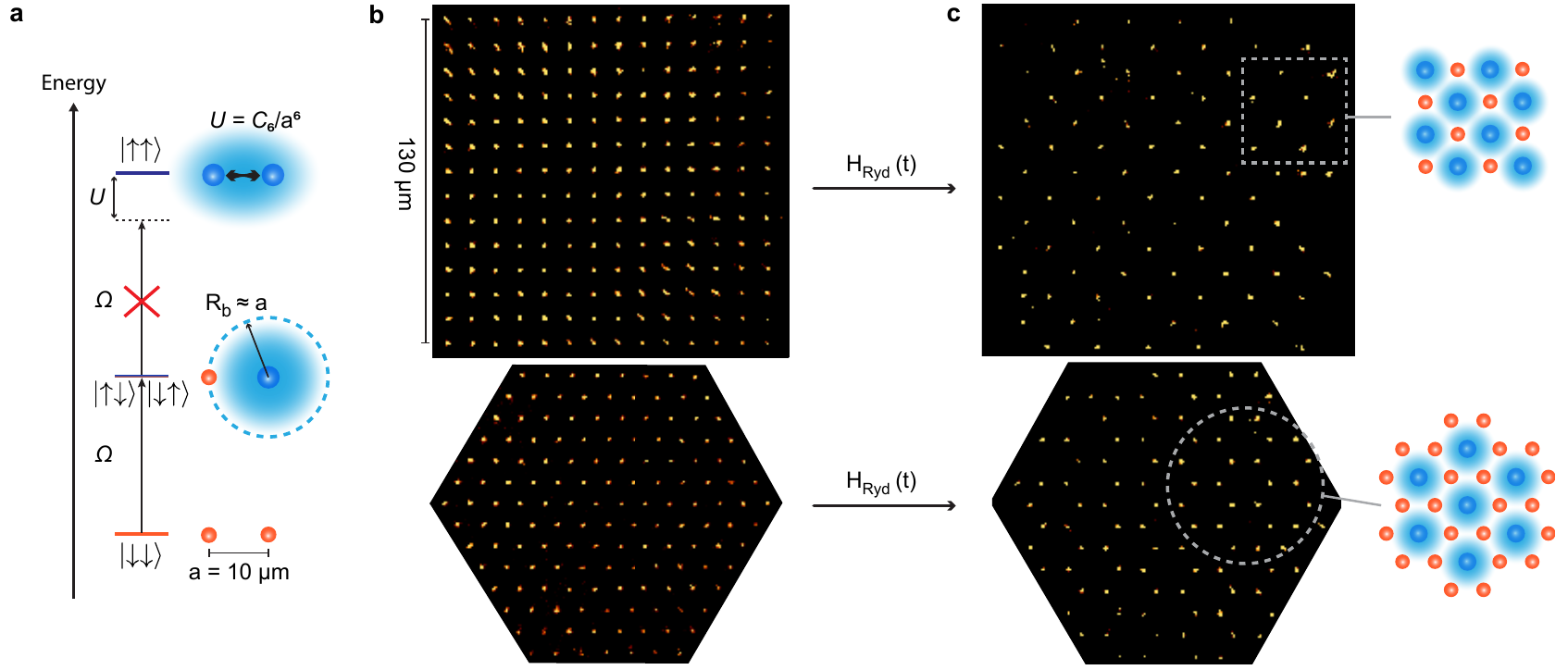}
	\caption{\textbf{Emergence of antiferromagnetic ordering from the Rydberg 
			blockade in square and triangular arrays.} 
		\textbf{a}, Illustration of the Rydberg blockade 
		with two atoms, whereby the strong interactions prevent the simultaneous 
		excitation of two atoms from the ground state (red circles) to the Rydberg state (blue circles) within the 
		Rydberg blockade radius $R_{\rm{b}}$ at which 
		$U=\hbar\Omega$. 
		\textbf{b},\textbf{c} Single-shot fluorescence images of ground state ($\ket{\downarrow}$) 
		atoms in a ${14\times14}$ square array and a 147-atom triangular 
		array with an atomic separation of $a=10~\mu$m. \textbf{b}, Initial PM states and \textbf{c}, nearly perfect AF ordering.
	}
	\label{fig1}
\end{figure*}

Previous works demonstrated the potential of Rydberg-based quantum simulators with up to a few tens of 
atoms\cite{Bernien2017,Keesling2019,Lienhard2018}, including 
high-fidelity manipulations \cite{Levine2019,Omran2019,Madjarov2020}. 
In particular, the transverse field Ising model (TFI) 
has been studied in 1D with up to 51 atoms\cite{Schauss2015,Bernien2017,Keesling2019}, 
in 2D square arrays, but with a limited degree of 
coherence\cite{Lienhard2018,GuardadoSanchez2018} making it difficult to observe 
genuine quantum features,
and recently in 3D with 22 atoms\cite{Song2020}. 
Here we implement the TFI in 2D, combining much larger atom numbers (up to $\sim 200$) 
and a high degree of coherence. In our implementation,
we explore two geometries which exhibit 
qualitatively different phase diagrams:
the bipartite square lattice and the geometrically frustrated triangular lattice\cite{Wannier1950}. 
On the square lattice, we prepare the N\'eel state 
characteristic of antiferromagnets with unprecedented probability.  On the triangular lattice, we observe 
for the first time, the creation of two distinct antiferromagnetic 
orders. The large number of atoms involved makes direct comparison between experimental results 
and the best numerical simulations extremely challenging. 
To validate the dynamics of our simulator we have pushed matrix product state simulations to their
limit and are able to simulate the dynamics of up to 100 atoms in 2D. We obtain an 
impressive agreement between the simulation and the experiment up to this number, 
which is one of the largest for which a direct comparison has been performed.  
Finally, by comparing the experiment to classical Monte Carlo 
calculations we demonstrate that our results cannot be 
reproduced by a classical equilibrium distribution at the same mean energy, and that the experiment features
an enhanced probability of finding classical ground states.

For arrays of atoms coupled by the (repulsive) van der Waals 
interaction, when excited to Rydberg states, the Hamiltonian of the TFI model is:
\begin{equation}
H_{\rm{Ryd}}=\sum_{i < j} U_{ij} n_i n_j + \frac{\hbar\Omega}{2}\sum_i \sigma^x_i - \hbar\delta\sum_i n_i,
\label{eq:tfiRyd}
\end{equation}
where the Rydberg and ground states are mapped onto the (pseudo-) spin states 
$\ket{\uparrow}$ and $\ket{\downarrow}$, respectively.
Here ${U_{ij} =C_6/{r_{ij}^6}}$ is the van der Waals interaction, $r_{ij}$ 
is the distance between atoms $i$ and $j$, 
$n_i=\ket{\uparrow}\bra{\uparrow}_i=(1+\sigma_i^z)/2$, and $\sigma_i$ are the usual Pauli matrices. 
The two spin states are coupled via a laser field with a Rabi frequency $\Omega$ and 
a detuning $\delta$, which act as transverse and longitudinal fields, respectively. 
Antiferromagnetic (AF) ordering in the system appears as a consequence of the strong interactions 
characterised by the Rydberg blockade radius $R_{\rm{b}}$, as illustrated in \figref{fig1}a\cite{Jaksch2000}. 
The type of antiferromagnetic ordering depends on the geometry of the array and the 
Hamiltonian parameters. 

\begin{figure*}[t]
	\includegraphics{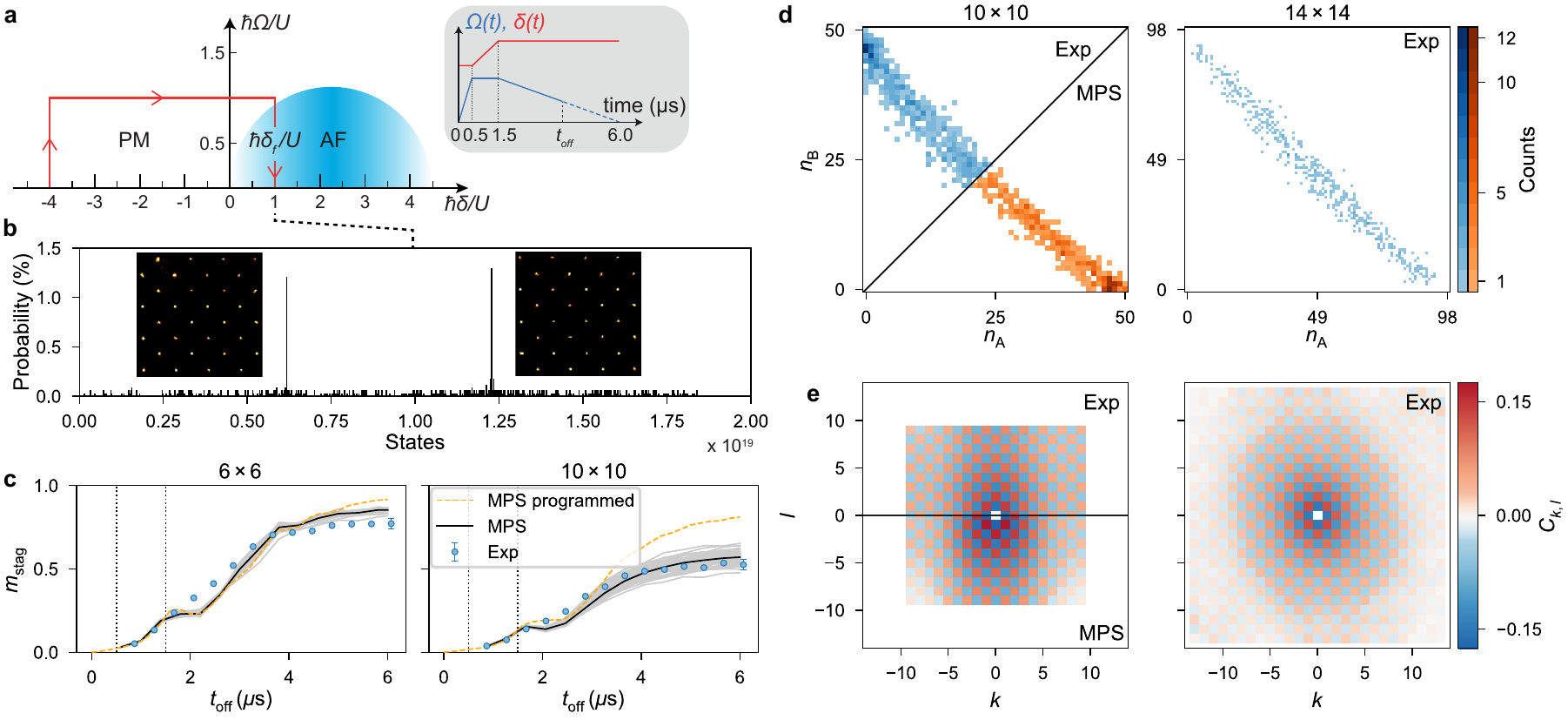}
	\caption{\textbf{The Ising model on a square lattice.} 
		\textbf{a}, Sketched bulk phase diagram for the square lattice.
		The inset shows the sweep shape with $t_{\rm off}$ the
		switch-off time of the excitation laser. The corresponding trajectory in the phase diagram is shown as a red arrow.
		\textbf{b}, State histogram for the ${8\times8}$ array at the end of the sweep. 
		The insets show fluorescence images of the two perfect AF states,
		which are obtained with 2.5\% probability. \textbf{c}, Growth of the staggered magnetisation during the sweep for the $6 \times 6$ and $10 \times 10$ arrays. The
		blue circles are experimental results with standard errors on the mean smaller than the markers size. The error bar on the final point is indicative of the long-term stability of the machine (see Sec.~\ref{SM:stability}). MPS simulations without (dashed line) and with (grey lines) experimental imperfections for which 50 ($6\times6$), 77 ($10\times10$) disorder instances are shown, with their average shown in black. 
		The vertical dotted lines correspond to the turning points in the sweep.
		\textbf{d}, Final staggered magnetisation histograms for the ${10\times10}$ and 
		${14\times14}$ arrays and \textbf{e}, corresponding 
		correlation maps, with MPS results shown in the lower half for the 
		${10\times10}$ array.}
	\label{fig2}
\end{figure*}

We create defect-free square and triangular arrays  
of respectively up to 196 and 147 $^{87}$Rb atoms using an 
optimised atom-by-atom assembly protocol\cite{Schymik2020} (see \figref{fig1}b). 
We define $\ket{\downarrow} = \ket{5S_{1/2}, F=2, m_{\rm{F}}=2}$
and $\ket{\uparrow} = \ket{75S_{1/2},m_{\rm{J}}=1/2}$, which are coupled via 
the intermediate state $\ket{6P_{3/2}, F=3, m_{\rm{F}}=3}$ with two counter-propagating 
laser beams of wavelength 420~nm and 1013~nm\cite{Levine2018} (see Sec.~\ref{SM:expSetup}). 
We achieve a single atom excitation probability of $99\%$ and a coherence time $\sim20$ 
times longer than in our previous work\cite{Lienhard2018} (see Sec.~\ref{SM:rabi}). 
We use arrays with atomic spacing $a=10~\mu$m, leading to a nearest-neighbour interaction
of $U/h\simeq 1.95{\;\rm MHz}$.

To probe the phase diagram of $H_{\rm{Ryd}}$
we
sweep $\Omega(t)$ and $\delta(t)$, 
and transfer the system from its initial paramagnetic (PM) ground state 
$\ket{\downarrow\downarrow\dots\downarrow}$ into the AF phase. 
A quantum phase transition (QPT) separates these two phases. Ideally, 
one would adiabatically drive the system such that it remains in the instantaneous ground state. 
However,  the energy gap at the QPT decreases with the atom number 
$N$, ($\sim 1/\sqrt{N}$ on a square lattice and $\sim\exp(-\alpha N)$ 
for the triangular lattice\cite{Janke1997,Laumann2012}). This leads to time scales which are 
experimentally impractical due to decoherence effects. 
Hence, we choose sweep times ($\sim 6~\mu$s) which are short enough to avoid sizeable decoherence but sufficiently long to quasi-adiabatically probe the phase diagram (see Sec.~\ref{SM:benchmarking4x4}).
We record fluorescence images of the atoms remaining in $\ket{\downarrow}$. Single-shot images, 
showing prepared, nearly perfect AF ordering on the square and triangular arrays, are shown in \figref{fig1}c.
For the results presented here, we typically repeat the sequence 1000 times. 

We first focus on the square lattice,using arrays of size ${N=L\times L}$, with even $L$ so that  
the two N\'eel states have the same energy. 
In \figref{fig2}a we 
sketch the (bulk) phase diagram. In the case of the van der Waals interaction implemented here, 
the AF phase region is expected to extend up to
$\hbar \Omega_{\rm c} \approx 1.25\ U$ at $\hbar \delta \approx 4.66\ U/2$ (Ref.~\citenum{Fey2019}).
More complex phases\cite{Samajdar2020} appear at the lower and upper 
boundaries of $\hbar \delta/U$ in the AF region.  
The applied sweeps are shown in \figref{fig2}a, 
with the QPT being crossed during the ramp down of $\Omega(t)$. 
Figure~\ref{fig2}b presents an experimental  histogram of the states recorded 
at the end of the sweep for the ${8\times8}$ array. 
Remarkably, out of $2^{64}\simeq2\times10^{19}$ possible states, 
we obtain a perfectly ordered state with a probability of $\sim$2.5\%, 
as can be seen by the two prominent peaks. 
The fluorescence images show the two corresponding N\'eel states.
To characterise the magnetic ordering of the states prepared during the sweep, 
we measure the order parameter, which is the normalised staggered 
magnetisation ${m_{\rm{stag}}=\langle|n_{\rm{A}}-n_{\rm{B}}|\rangle/(N/2)}$, 
giving the difference in the number of excitations on each sublattice $(\rm{A,B})$,
averaged over many realisations. 
The two perfect AF states correspond to  
one of the two sublattices being fully excited, such that $m_{\rm{stag}}=1$. 
We access the dynamics of the system during the sweep by rapidly 
turning off the excitation laser at different times $t_{\rm{off}}$ (see \figref{fig2}a).
Fig.~\ref{fig2}c presents the evolution of $m_{\rm{stag}}$ for the 
${6\times6}$ and ${10\times10}$ arrays, using the same sweep. Over the first $1.5~\mu$s of 
the sweep the system is in the PM phase,
where fluctuations lead to small but finite $m_{\rm stag}\sim1/\sqrt{N}$. 
We then observe the growth of $m_{\rm stag}$ during the drive 
of the system from the PM to the AF phase.

To benchmark our platform, we perform a systematic comparison of the 
dynamics with matrix product state (MPS) numerical simulations (see Sec.~\ref{SM:MPS}). 
We consider both the programmed 
and the real parameters, the latter of which include 
independently calibrated experimental imperfections, 
with the exception of decoherence effects (see Sec.~\ref{SM:imp}). 
For the ${6\times6}$ array, we observe a good agreement between 
the experimental results and the MPS simulations, for both 
situations. For the ${10\times10}$ array, the experiment and the 
real MPS simulations also agree well. The difference between the programmed and 
the real MPS simulations highlights that the imperfections have a more severe impact on larger systems. 
Additionally, the reduced final value of $m_{\rm stag}$ for the programmed MPS on the $10\times10$ array indicates that as the system size grows, adiabaticity is indeed harder to achieve.

We now characterise the final state obtained at the end of the sweep ($\Omega=0$). 
Firstly, we visualise the shot-wise contributions to 
$m_{\rm{stag}}$ using a 2D histogram of the probability 
$P(n_{\rm{A}},n_{\rm{B}})$ of the $\ket{\uparrow}$ populations 
$n_{\rm{A}}$ and $n_{\rm{B}}$ of the two sublattices A, B. 
Here, the two N\'eel states appear as points at ($N/2,0$) and ($0,N/2$). 
The results are plotted in Fig.~\ref{fig2}d for the ${10\times10}$ and ${14\times14}$ arrays. 
For both systems we observe the presence of points along the 
diagonal highlighting that the average Rydberg density is $\sim 50\%$. 
For the ${10\times10}$ we observe a conglomeration of points 
around the two corners belonging to the N\'eel states.
Because of the imperfections and the scaling of the energy gap, 
the state preparation becomes more challenging with increasing system size. 
The elongated histogram for the $14 \times 14$ array 
demonstrates that, remarkably, we prepare strongly AF ordered states 
($m_{\rm stag} \sim 0.4$), even for such large systems. This is also
evident in the fluorescence image in \figref{fig1}c, containing 184 atoms (out of 196) obeying AF ordering.
For a comparison with simulations, we have devised an algorithm 
to stochastically sample the MPS wavefunction, thereby obtaining 
snapshots as in the experiment (see Sec.~\ref{SM:mpsSampling}).
The lower half of \figref{fig2}d shows the so-obtained histogram for the 
$10\times10$ lattice, which matches the experiment very well.
For even larger atom numbers, accurate MPS simulations become intractable.

Secondly, we compute the connected spin-spin correlation function defined as
\begin{equation}
C_{k,l} = \frac{1}{N_{k,l}}\sum_{i,j}{\langle n_in_j\rangle - \langle n_i\rangle \langle n_j\rangle},
\label{eq:ckl}
\end{equation}
where the sum runs over all pairs of atoms $i,j$ separated by $k\textbf{e}_1+l\textbf{e}_2$, with $\textbf{e}_{1(2)}$ denoting the two vectors of the underlying lattice, and $N_{k,l}$ being the number of such pairs.
Figure~\ref{fig2}e shows the $C_{k,l}$ correlation maps corresponding to the 
$m_{\rm{stag}}$ histograms shown in \figref{fig2}d. The plots display the alternation 
of correlation and anti-correlation, expected for AF ordering, 
whose values would be $\pm1/4$ for the N\'eel state. 
The spatial decay of the correlations is well described by correlation 
lengths of $\xi\simeq7a$ and $5.5a$ for the two system sizes respectively, 
showing that the sweeps produce highly AF ordered states\footnote{The residual 
	anisotropy observed is due to the finite size of the excitation beams, 
	which is comparable to the width of the array ($130~\mu$m).}. 
Again we observe very good agreement between experimental 
and real MPS results for the ${10\times10}$ array, confirming that 
the simulations capture well the experimental conditions 
(for a real-time analysis of the correlations during the sweep, see Sec.~\ref{SM:AFM}).

\begin{figure}[t]
	\includegraphics[width = 8.5cm]{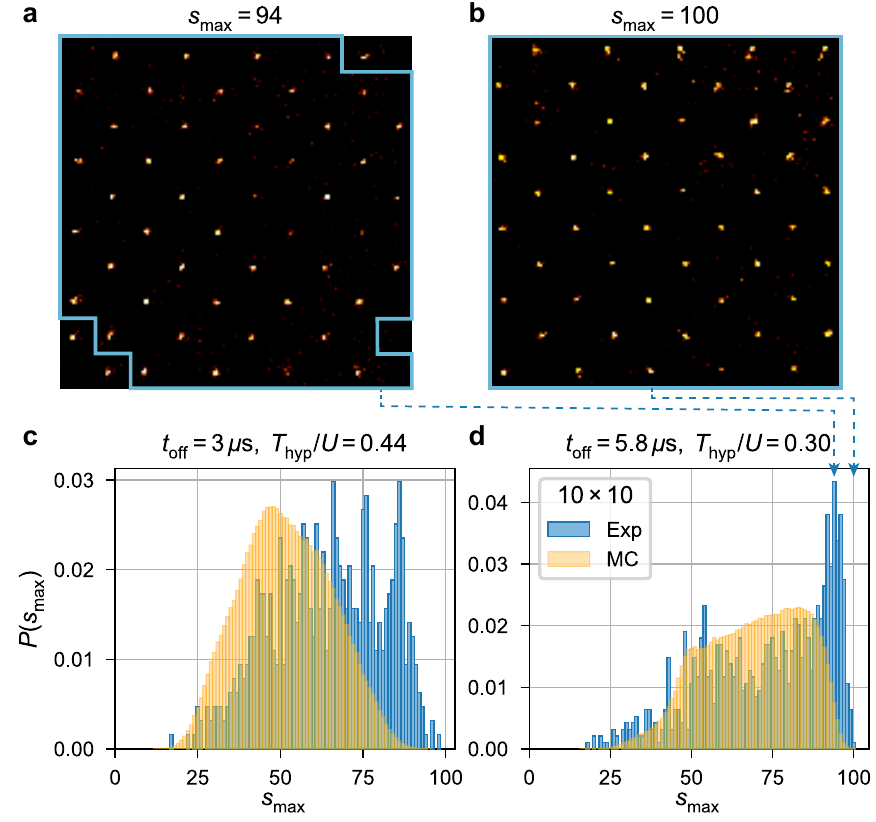}
	\caption{\textbf{Quantum real-time evolution versus classical equilibrium:}
		\textbf{a, b}, Fluorescence images on a ${10\times10}$ array illustrating how we extract the largest AF domains comprising of $s_{\rm max}$ sites, indicated by the blue boundaries.
		\textbf{c, d}, Distributions of $s_{\rm max}$ 
		during \textbf{c} and at the end of \textbf{d} the sweep (blue), 
		compared to the classical equilibrium result (yellow)
		with the corresponding hypothetical temperature $T_{\rm hyp}$.}
	\label{fig3}
\end{figure}

To further quantify the AF ordering, we analyse 
the distribution of antiferromagnetic cluster sizes~\cite{Stoli1979}. 
For each run of the experiment, 
we decompose the snapshot into individual clusters obeying local AF ordering
(see examples in \figref{fig3}a,b). 
We count the number of atoms 
inside each individual cluster, and record the largest size, $s_{\rm max}$. 
From the full set of snapshots we reconstruct the probability distribution 
$P(s_{\rm max})$. For a perfectly AF-ordered state, this distribution presents 
as a single peak of unit probability at ${s_\text{max} = N}$, while imperfect 
ordering shows up as a distribution broadened towards smaller 
$s_{\rm max}$. In \figref{fig3}c,d we show $P(s_{\rm max})$ at two 
instants during the sweep for the  ${10\times10}$ array (blue bars). 
Even at intermediate times, shortly after entering the AF region (\figref{fig3}c), 
we observe the presence of significant AF ordering, as the distribution 
is centred around large values of $s_\text{max}$. 
At the end of the sweep (\figref{fig3}d), N\'eel states are observed, 
as shown in \figref{fig3}b, and more than 27\% of the shots feature AF states 
with clusters missing at most 10 sites.

The fact that we obtain a distribution of final states raises the question 
whether the system has thermalised during the finite duration of the sweep~\cite{Bernien2017,Kim2018}.
To answer this question, we compare the observed distribution $P(s_{\rm max})$ to the corresponding distribution obtained from a classical equilibrium setup with a hypothetical temperature $T_{\rm hyp}$. 
We focus on a classical description for two reasons: 
i) the classical energy is the one accessible in the experiment 
and ii) at the end of the sweep, $\Omega=0$ and the quantum and classical 
statistical mechanics descriptions coincide.

To determine $T_{\rm hyp}$ we match the classical Ising energy $E_{\rm class}(t_{\rm off})$ 
of the experimental system with
$E^{\rm MC}_{\rm class}(T)$ from the corresponding classical statistical 
mechanics system for a given temperature $T$ estimated from a 
Monte Carlo (MC) sampling 
\footnote{One 
could also use other observables to match a 
hypothetical temperature, but since temperature is the variable 
conjugate to the energy in thermodynamics, it is the most natural choice.}. 
We refer to Sec.~\ref{SM:T_hyp} for a thorough discussion of $T_{\rm hyp}$ during the sweep.
In \figref{fig3}c,d, we show $P(s_{\rm max})$ for the corresponding 
classical equilibrium distributions (orange bars), and observe that they 
are centered at a significantly smaller $s_\text{max}$, and do not reproduce 
the distribution of the experimental results. In particular, the probability of 
finding perfectly ordered states is much higher in the experiment.
An equivalent analysis of the MPS real-time evolution shows similar 
features (see Sec.~\ref{SM:QvsC}). 
Our analysis therefore reveals that despite residual imperfections, the experiment 
does not thermalise during the state preparation protocol and 
can only be consistently reproduced by a unitary quantum mechanical real-time description. 
Furthermore, the enhanced probability of finding the targeted classical states 
is promising for future applications of the Rydberg platform, e.g., 
as a quantum annealer to solve optimisation problems of various 
types\cite{Pichler2018,Henriet2020a,Serret2020}.

\begin{figure*}[t]
	\includegraphics{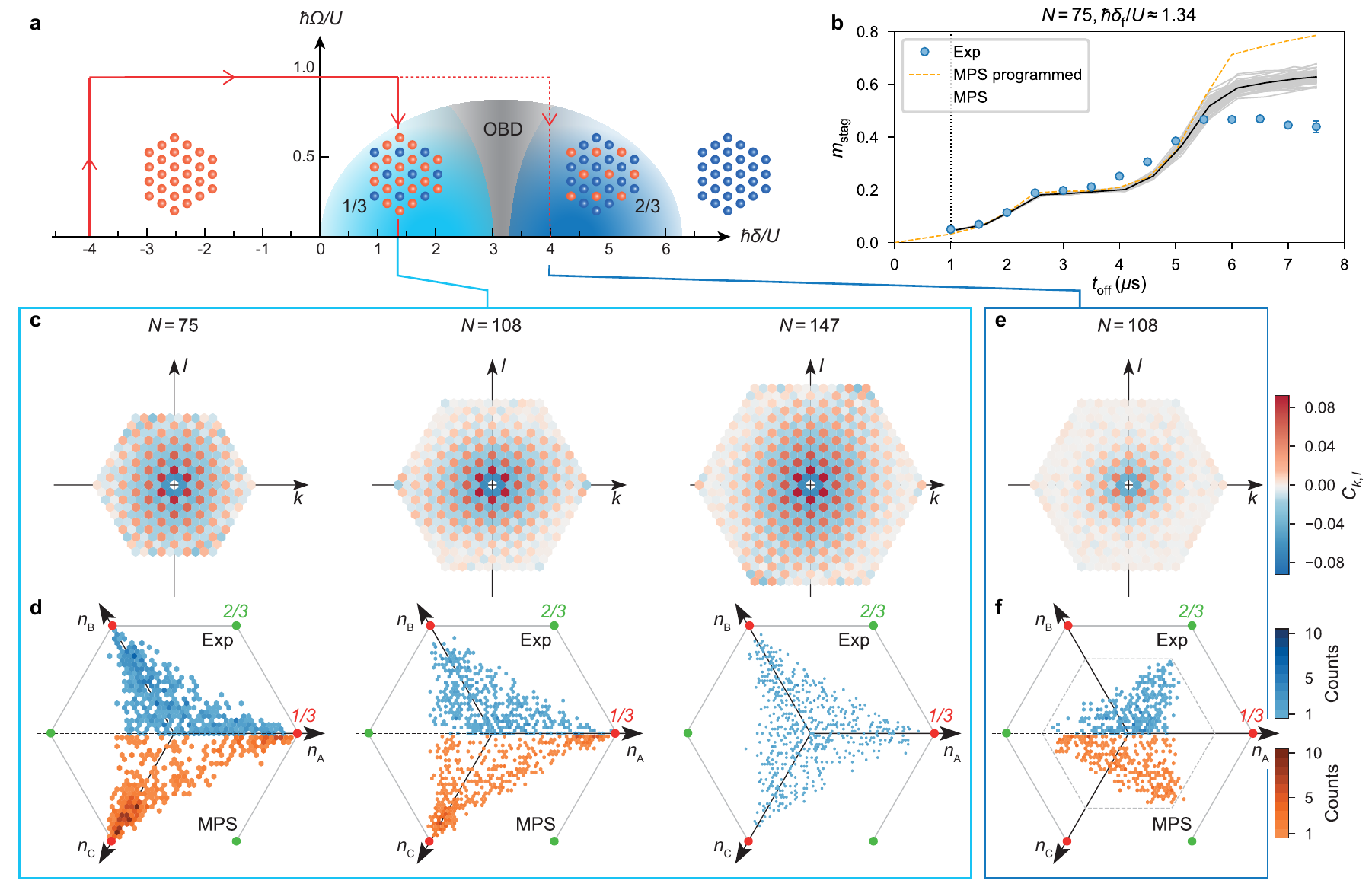}
	\caption{\textbf{Antiferromagnetic ordering on a triangular array.}
		\textbf{a}, Sketched bulk phase diagram for the triangular lattice.
		The red arrows show the sweeps used to prepare the 
		phases at filling $1/3$ and $2/3$. 
		\textbf{b}, Growth of the staggered magnetisation during the sweep for a 75-atom array. The
		blue circles are experimental results with standard errors on the mean smaller than the markers size. The error bar on the final point is indicative of the long-term stability of the machine (see Sec.~\ref{SM:stability}). MPS simulations without (dashed line) and with (grey lines) experimental imperfections for which 50 disorder instances are shown, with their average shown in black. 
		The vertical dotted lines correspond to the turning points in the sweep.
		\textbf{c}, Final experimental correlation maps and \textbf{d}, corresponding staggered 
		magnetisation histograms plotted in the complex plane for the 75, 
		108 and 147 site triangular arrays for the sweep preparing the $1/3$ phase.
		For the 75 and 108 site arrays the lower half of the histograms show the analogous MPS results.
		\textbf{e}, Final experimental correlation map and \textbf{g}, corresponding staggered 
		magnetisation histogram for the 108 site triangular array for the sweep preparing the $2/3$ phase.
		The lower half in \textbf{f} shows corresponding MPS results. 
	}
	\label{fig4}
\end{figure*}

Having explored the square lattice we now consider the more complex triangular array. Here, the TFI model
features a richer phase diagram, with prominent ordered phases at $1/3$ and $2/3$ Rydberg filling, 
as sketched in \figref{fig4}a.
The $1/3$ phase is the analogue of the AF 
ordering on the square lattice, where the Rydberg blockade prevents neighbouring 
sites from being excited simultaneously, leading to one of the three sublattices being filled with Rydberg excitations, illustrated in \figref{fig1}c. 
The $2/3$ phase is the ``particle-hole'' inverse of this, with two sublattices 
being fully excited and one sublattice containing ground state atoms.
In between these phase regions, at 1/2 filling, the classical Ising model ($\Omega=0$) 
is strongly frustrated for nearest-neighbour interactions, with 
an exponentially large (in $N$) ground state manifold \cite{Wannier1950}.
Finite $\Omega$ stabilises yet another ordered phase in a process called ``order by disorder'' (OBD)~\cite{Villain1980, Moessner2000, Moessner2001, Isakov2003, Fey2019, Koziol2019}.

To explore the triangular phase diagram, we consider hexagonal clusters of various sizes, 
built shell by shell around a central three-atom triangle (see \figref{fig1}b). 
We apply the sweeps shown in \figref{fig4}a for two different final detunings 
$\delta_\text{f}$, to create the $1/3$ and the $2/3$ phases. 
To quantify the state preparation process, we again measure the temporal dynamics of the order parameter, the normalised staggered magnetisation. For the triangular array, this is defined as 
$m_\text{stag} = \langle\left|n_{\rm{A}} + \E^{i 2\pi/3} n_{\rm{B}} + \E^{-i 2\pi/3} n_{\rm{C}}\right|\rangle/(N/3)$, 
where $n_{\rm A,B,C}$ is the Rydberg population on each of the three sublattices. 
We plot the results in \figref{fig4}b for the experiment and two types of MPS simulations (programmed and 
real) for a 75-atom array, with $\delta_\text{f}$ chosen to prepare the $1/3$ phase. 
We observe the growth of the AF ordering both in the experiment and the simulations, which agree very well during the first 5~$\mu$s of the sweep. After this the experimental results plateau at a lower value of $m_\mathrm{stag}$ than expected from the MPS. The inclusion of experimental imperfections decreases the final value of $m_{\rm stag}$, however, there is still a discrepancy with the experimental results.
A possible explanation could be the enhanced sensitivity of the QPT from 
paramagnetic to the $1/3$ AF phase (believed to be first-order\cite{Janke1997}) 
to the residual experimental imperfections 
not included in the MPS simulation. 
Confirming the origin of this effect will be the subject of future work.

To further characterise the prepared final states we consider $C_{k,l}$, defined similarly to Eq.~(\ref{eq:ckl}).
Here, the perfect AF state would have $C_{k,l}=+2/9$ and $-1/9$ for correlated and anti-correlated sites, respectively.
In \figref{fig4}c,e we show the final state correlation maps for the 1/3 (2/3) phases on atom arrays 
of 75, 108 and 147 sites (108 sites). 
We observe a pattern characteristic of three-sublattice ordered states,
throughout almost the entire bulk of our systems, with a correlation length 
$\xi \sim 3$---$3.7 a$ for the $1/3$ phase and $\xi \sim 2 a$ for the $2/3$ phase.
In \figref{fig4}d,f we plot the corresponding distributions of the complex order parameter. 
For perfectly ordered $1/3$ ($2/3$) systems, one would expect peaks 
at the three corners of the bounding hexagon, marked by red (green) dots. 
We observe correspondingly aligned triangular distributions with good 
agreement between the experimental and MPS results. 
There is a higher accumulation of points at the corners in 
the MPS results, which corresponds to a higher value of $m_\mathrm{stag}$. 
While the distributions spread almost fully to the corners in the $1/3$ 
phase results, for the $2/3$ phase the size of the triangle is visibly reduced.
This reduction 
is due to finite-size cluster effects: 
the boundary is filled with Rydberg excitations which
reduces the maximal possible extent of the distribution to the inner hexagon (dashed lines) 
(see Sec.~\ref{SM:tri2over3}).
To our knowledge, the above results are the first observations of the 
1/3 and 2/3 phases using a quantum many-body system. Despite the experimental 
imperfections, and the finite sweep duration, we are able to produce highly-ordered 
states on even the largest 147-atom array. This is highlighted in \figref{fig1}c showing a 
fluorescence image of the 1/3 phase on the 147-atom array with almost perfect AF ordering.
Finally, similar to the square array, we observe an enhanced probability 
of finding highly ordered states compared to a classical equilibrium system 
at the same energy, as revealed by experimental order parameter distributions 
which are centred at larger values (see details in Sec.~\ref{SM:MCvsExpTri}).

As a conclusion, we have probed the quantum dynamics of Ising magnets in square 
and triangular geometries, beyond situations which can be exactly 
simulated classically. 
We have validated the experimental results with comprehensive numerical 
simulations up to computationally feasible sizes. We have shown a 
high degree of coherence and control, over a large number 
of atoms. Combined, this demonstrates that our platform is now able to study quantum spin models in regimes 
beyond those accessible via numerical investigations.
We have also identified a potential advantage of 
Rydberg quantum simulators to prepare targeted classical states, 
compared to classical equilibrium systems. 
Natural extensions of this work include a thorough 
investigation of the dynamics of the 2D quantum phase transition, 
and further explorations of the effects of frustration, in particular the 
observation of the elusive OBD phase. 
Finally, our benchmark provides a roadmap for improving the platform even further
thus opening exciting prospects beyond quantum simulation, 
e.g. for optimisation\cite{Pichler2018,Serret2020}, 
quantum sensing\cite{Madjarov2019,Norcia2019} and quantum computing\cite{Saffman2010,Henriet2020b,Whitlock2020}.

\begin{acknowledgments}
	This project has received funding from the European Union's Horizon 2020 research and innovation program under grant agreement no. 817482 (PASQuanS).
	MS acknowledges support by the Austrian Science Fund (FWF) through Grant No.~P~31701 (ULMAC).
	DB acknowledges support from the Ramón y Cajal program (RYC2018-025348-I).
	KNS acknowledges support by the Studienstiftung des Deutschen Volkes.
	AAE and AML acknowledge support by the Austrian Science Fund (FWF) through Grant No.~I~4548.
	The computational results presented have been achieved in part using the Vienna Scientific Cluster (VSC) and the LEO HPC infrastructure of the University of Innsbruck.
\end{acknowledgments}

\section*{Author contributions}
\noindent
$^*$PS, MS, HJW, and AAE contributed equally to this work.
PS, HJW, DB, KS, and VL carried out the experiments. MS, AAE, L-PH and TCL performed the simulations. All authors contributed to the data analysis, progression of the project, and on both the experimental and theoretical side. All authors contributed to the writing of the manuscript. Correspondence and requests for materials should be addressed to Pascal Scholl.

\section*{Ethics Declaration}
The authors declare no competing interests.

\bibliography{Bib}

\begin{thebibliography}{60}%
\makeatletter
\providecommand \@ifxundefined [1]{%
 \@ifx{#1\undefined}
}%
\providecommand \@ifnum [1]{%
 \ifnum #1\expandafter \@firstoftwo
 \else \expandafter \@secondoftwo
 \fi
}%
\providecommand \@ifx [1]{%
 \ifx #1\expandafter \@firstoftwo
 \else \expandafter \@secondoftwo
 \fi
}%
\providecommand \natexlab [1]{#1}%
\providecommand \enquote  [1]{``#1''}%
\providecommand \bibnamefont  [1]{#1}%
\providecommand \bibfnamefont [1]{#1}%
\providecommand \citenamefont [1]{#1}%
\providecommand \href@noop [0]{\@secondoftwo}%
\providecommand \href [0]{\begingroup \@sanitize@url \@href}%
\providecommand \@href[1]{\@@startlink{#1}\@@href}%
\providecommand \@@href[1]{\endgroup#1\@@endlink}%
\providecommand \@sanitize@url [0]{\catcode `\\12\catcode `\$12\catcode
  `\&12\catcode `\#12\catcode `\^12\catcode `\_12\catcode `\%12\relax}%
\providecommand \@@startlink[1]{}%
\providecommand \@@endlink[0]{}%
\providecommand \url  [0]{\begingroup\@sanitize@url \@url }%
\providecommand \@url [1]{\endgroup\@href {#1}{\urlprefix }}%
\providecommand \urlprefix  [0]{URL }%
\providecommand \Eprint [0]{\href }%
\providecommand \doibase [0]{http://dx.doi.org/}%
\providecommand \selectlanguage [0]{\@gobble}%
\providecommand \bibinfo  [0]{\@secondoftwo}%
\providecommand \bibfield  [0]{\@secondoftwo}%
\providecommand \translation [1]{[#1]}%
\providecommand \BibitemOpen [0]{}%
\providecommand \bibitemStop [0]{}%
\providecommand \bibitemNoStop [0]{.\EOS\space}%
\providecommand \EOS [0]{\spacefactor3000\relax}%
\providecommand \BibitemShut  [1]{\csname bibitem#1\endcsname}%
\let\auto@bib@innerbib\@empty
\bibitem [{\citenamefont {Georgescu}\ \emph {et~al.}(2014)\citenamefont
  {Georgescu}, \citenamefont {Ashhab},\ and\ \citenamefont
  {Nori}}]{GeorgescuRMP}%
  \BibitemOpen
  \bibfield  {author} {\bibinfo {author} {\bibfnamefont {I.~M.}\ \bibnamefont
  {Georgescu}}, \bibinfo {author} {\bibfnamefont {S.}~\bibnamefont {Ashhab}}, \
  and\ \bibinfo {author} {\bibfnamefont {F.}~\bibnamefont {Nori}},\ }\emph
  {Quantum simulation},\ \href {\doibase 10.1103/RevModPhys.86.153} {\bibfield
  {journal} {\bibinfo  {journal} {Rev. Mod. Phys.}\ }\textbf {\bibinfo {volume}
  {86}},\ \bibinfo {pages} {153} (\bibinfo {year} {2014})}\BibitemShut
  {NoStop}%
\bibitem [{\citenamefont {Blatt}\ and\ \citenamefont {Roos}(2012)}]{Blatt2012}%
  \BibitemOpen
  \bibfield  {author} {\bibinfo {author} {\bibfnamefont {R.}~\bibnamefont
  {Blatt}}\ and\ \bibinfo {author} {\bibfnamefont {C.~F.}\ \bibnamefont
  {Roos}},\ }\emph {Quantum simulations with trapped ions},\ \href {\doibase
  10.1038/nphys2252} {\bibfield  {journal} {\bibinfo  {journal} {Nat. Phys.}\
  }\textbf {\bibinfo {volume} {8}},\ \bibinfo {pages} {277} (\bibinfo {year}
  {2012})}\BibitemShut {NoStop}%
\bibitem [{\citenamefont {Monroe}\ \emph {et~al.}(2020)\citenamefont {Monroe},
  \citenamefont {Campbell}, \citenamefont {Duan}, \citenamefont {Gong},
  \citenamefont {Gorshkov}, \citenamefont {Hess}, \citenamefont {Islam},
  \citenamefont {Kim}, \citenamefont {Linke}, \citenamefont {Pagano},
  \citenamefont {Richerme}, \citenamefont {Senko},\ and\ \citenamefont
  {Yao}}]{Monroe2020}%
  \BibitemOpen
  \bibfield  {author} {\bibinfo {author} {\bibfnamefont {C.}~\bibnamefont
  {Monroe}}, \bibinfo {author} {\bibfnamefont {W.~C.}\ \bibnamefont
  {Campbell}}, \bibinfo {author} {\bibfnamefont {L.-M.}\ \bibnamefont {Duan}},
  \bibinfo {author} {\bibfnamefont {Z.-X.}\ \bibnamefont {Gong}}, \bibinfo
  {author} {\bibfnamefont {A.~V.}\ \bibnamefont {Gorshkov}}, \bibinfo {author}
  {\bibfnamefont {P.}~\bibnamefont {Hess}}, \bibinfo {author} {\bibfnamefont
  {R.}~\bibnamefont {Islam}}, \bibinfo {author} {\bibfnamefont
  {K.}~\bibnamefont {Kim}}, \bibinfo {author} {\bibfnamefont {N.}~\bibnamefont
  {Linke}}, \bibinfo {author} {\bibfnamefont {G.}~\bibnamefont {Pagano}},
  \bibinfo {author} {\bibfnamefont {P.}~\bibnamefont {Richerme}}, \bibinfo
  {author} {\bibfnamefont {C.}~\bibnamefont {Senko}}, \ and\ \bibinfo {author}
  {\bibfnamefont {N.~Y.}\ \bibnamefont {Yao}},\ }\emph {Programmable Quantum
  Simulations of Spin Systems with Trapped Ions},\ \href
  {https://arxiv.org/abs/1912.07845} {\bibfield  {journal} {\bibinfo  {journal}
  {arXiv:1912.07845}\ } (\bibinfo {year} {2020})}\BibitemShut {NoStop}%
\bibitem [{\citenamefont {Gärttner}\ \emph {et~al.}(2017)\citenamefont
  {Gärttner}, \citenamefont {Bohnet}, \citenamefont {Safavi-Naini},
  \citenamefont {Wall}, \citenamefont {Bollinger},\ and\ \citenamefont
  {Rey}}]{Garttner2017}%
  \BibitemOpen
  \bibfield  {author} {\bibinfo {author} {\bibfnamefont {M.}~\bibnamefont
  {Gärttner}}, \bibinfo {author} {\bibfnamefont {J.~G.}\ \bibnamefont
  {Bohnet}}, \bibinfo {author} {\bibfnamefont {A.}~\bibnamefont
  {Safavi-Naini}}, \bibinfo {author} {\bibfnamefont {M.~L.}\ \bibnamefont
  {Wall}}, \bibinfo {author} {\bibfnamefont {J.~J.}\ \bibnamefont {Bollinger}},
  \ and\ \bibinfo {author} {\bibfnamefont {A.~M.}\ \bibnamefont {Rey}},\ }\emph
  {Measuring out-of-time-order correlations and multiple quantum spectra in a
  trapped-ion quantum magnet},\ \href {\doibase 10.1038/nphys4119} {\bibfield
  {journal} {\bibinfo  {journal} {Nat. Phys.}\ }\textbf {\bibinfo {volume}
  {13}},\ \bibinfo {pages} {781} (\bibinfo {year} {2017})}\BibitemShut
  {NoStop}%
\bibitem [{\citenamefont {Song}\ \emph {et~al.}(2017)\citenamefont {Song},
  \citenamefont {Xu}, \citenamefont {Liu}, \citenamefont {Yang}, \citenamefont
  {Zheng}, \citenamefont {Deng}, \citenamefont {Xie}, \citenamefont {Huang},
  \citenamefont {Guo}, \citenamefont {Zhang}, \citenamefont {Zhang},
  \citenamefont {Xu}, \citenamefont {Zheng}, \citenamefont {Zhu}, \citenamefont
  {Wang}, \citenamefont {Chen}, \citenamefont {Lu}, \citenamefont {Han},\ and\
  \citenamefont {Pan}}]{Song2017}%
  \BibitemOpen
  \bibfield  {author} {\bibinfo {author} {\bibfnamefont {C.}~\bibnamefont
  {Song}}, \bibinfo {author} {\bibfnamefont {K.}~\bibnamefont {Xu}}, \bibinfo
  {author} {\bibfnamefont {W.}~\bibnamefont {Liu}}, \bibinfo {author}
  {\bibfnamefont {C.-p.}\ \bibnamefont {Yang}}, \bibinfo {author}
  {\bibfnamefont {S.-B.}\ \bibnamefont {Zheng}}, \bibinfo {author}
  {\bibfnamefont {H.}~\bibnamefont {Deng}}, \bibinfo {author} {\bibfnamefont
  {Q.}~\bibnamefont {Xie}}, \bibinfo {author} {\bibfnamefont {K.}~\bibnamefont
  {Huang}}, \bibinfo {author} {\bibfnamefont {Q.}~\bibnamefont {Guo}}, \bibinfo
  {author} {\bibfnamefont {L.}~\bibnamefont {Zhang}}, \bibinfo {author}
  {\bibfnamefont {P.}~\bibnamefont {Zhang}}, \bibinfo {author} {\bibfnamefont
  {D.}~\bibnamefont {Xu}}, \bibinfo {author} {\bibfnamefont {D.}~\bibnamefont
  {Zheng}}, \bibinfo {author} {\bibfnamefont {X.}~\bibnamefont {Zhu}}, \bibinfo
  {author} {\bibfnamefont {H.}~\bibnamefont {Wang}}, \bibinfo {author}
  {\bibfnamefont {Y.-A.}\ \bibnamefont {Chen}}, \bibinfo {author}
  {\bibfnamefont {C.-Y.}\ \bibnamefont {Lu}}, \bibinfo {author} {\bibfnamefont
  {S.}~\bibnamefont {Han}}, \ and\ \bibinfo {author} {\bibfnamefont {J.-W.}\
  \bibnamefont {Pan}},\ }\emph {10-Qubit Entanglement and Parallel Logic
  Operations with a Superconducting Circuit},\ \href {\doibase
  10.1103/PhysRevLett.119.180511} {\bibfield  {journal} {\bibinfo  {journal}
  {Phys. Rev. Lett.}\ }\textbf {\bibinfo {volume} {119}},\ \bibinfo {pages}
  {180511} (\bibinfo {year} {2017})}\BibitemShut {NoStop}%
\bibitem [{\citenamefont {King}\ \emph {et~al.}(2018)\citenamefont {King},
  \citenamefont {Carrasquilla}, \citenamefont {Raymond}, \citenamefont
  {Ozfidan}, \citenamefont {Andriyash}, \citenamefont {Berkley}, \citenamefont
  {Reis}, \citenamefont {Lanting}, \citenamefont {Harris}, \citenamefont
  {Altomare}, \citenamefont {Boothby}, \citenamefont {Bunyk}, \citenamefont
  {Enderud}, \citenamefont {Fréchette}, \citenamefont {Hoskinson},
  \citenamefont {Ladizinsky}, \citenamefont {Oh}, \citenamefont
  {Poulin-Lamarre}, \citenamefont {Rich}, \citenamefont {Sato}, \citenamefont
  {Smirnov}, \citenamefont {Swenson}, \citenamefont {Volkmann}, \citenamefont
  {Whittaker}, \citenamefont {Yao}, \citenamefont {Ladizinsky}, \citenamefont
  {Johnson}, \citenamefont {Hilton},\ and\ \citenamefont {Amin}}]{King2018}%
  \BibitemOpen
  \bibfield  {author} {\bibinfo {author} {\bibfnamefont {A.~D.}\ \bibnamefont
  {King}}, \bibinfo {author} {\bibfnamefont {J.}~\bibnamefont {Carrasquilla}},
  \bibinfo {author} {\bibfnamefont {J.}~\bibnamefont {Raymond}}, \bibinfo
  {author} {\bibfnamefont {I.}~\bibnamefont {Ozfidan}}, \bibinfo {author}
  {\bibfnamefont {E.}~\bibnamefont {Andriyash}}, \bibinfo {author}
  {\bibfnamefont {A.}~\bibnamefont {Berkley}}, \bibinfo {author} {\bibfnamefont
  {M.}~\bibnamefont {Reis}}, \bibinfo {author} {\bibfnamefont {T.}~\bibnamefont
  {Lanting}}, \bibinfo {author} {\bibfnamefont {R.}~\bibnamefont {Harris}},
  \bibinfo {author} {\bibfnamefont {F.}~\bibnamefont {Altomare}}, \bibinfo
  {author} {\bibfnamefont {K.}~\bibnamefont {Boothby}}, \bibinfo {author}
  {\bibfnamefont {P.~I.}\ \bibnamefont {Bunyk}}, \bibinfo {author}
  {\bibfnamefont {C.}~\bibnamefont {Enderud}}, \bibinfo {author} {\bibfnamefont
  {A.}~\bibnamefont {Fréchette}}, \bibinfo {author} {\bibfnamefont
  {E.}~\bibnamefont {Hoskinson}}, \bibinfo {author} {\bibfnamefont
  {N.}~\bibnamefont {Ladizinsky}}, \bibinfo {author} {\bibfnamefont
  {T.}~\bibnamefont {Oh}}, \bibinfo {author} {\bibfnamefont {G.}~\bibnamefont
  {Poulin-Lamarre}}, \bibinfo {author} {\bibfnamefont {C.}~\bibnamefont
  {Rich}}, \bibinfo {author} {\bibfnamefont {Y.}~\bibnamefont {Sato}}, \bibinfo
  {author} {\bibfnamefont {A.~Y.}\ \bibnamefont {Smirnov}}, \bibinfo {author}
  {\bibfnamefont {L.~J.}\ \bibnamefont {Swenson}}, \bibinfo {author}
  {\bibfnamefont {M.~H.}\ \bibnamefont {Volkmann}}, \bibinfo {author}
  {\bibfnamefont {J.}~\bibnamefont {Whittaker}}, \bibinfo {author}
  {\bibfnamefont {J.}~\bibnamefont {Yao}}, \bibinfo {author} {\bibfnamefont
  {E.}~\bibnamefont {Ladizinsky}}, \bibinfo {author} {\bibfnamefont {M.~W.}\
  \bibnamefont {Johnson}}, \bibinfo {author} {\bibfnamefont {J.}~\bibnamefont
  {Hilton}}, \ and\ \bibinfo {author} {\bibfnamefont {M.~H.}\ \bibnamefont
  {Amin}},\ }\emph {{ Observation of topological phenomena in a programmable
  lattice of 1,800 qubits}},\ \href {\doibase 10.1038/s41586-018-0410-x}
  {\bibfield  {journal} {\bibinfo  {journal} {Nature}\ }\textbf {\bibinfo
  {volume} {560}},\ \bibinfo {pages} {456} (\bibinfo {year}
  {2018})}\BibitemShut {NoStop}%
\bibitem [{\citenamefont {Kjaergaard}\ \emph {et~al.}(2020)\citenamefont
  {Kjaergaard}, \citenamefont {Schwartz}, \citenamefont {Braumüller},
  \citenamefont {Krantz}, \citenamefont {Wang}, \citenamefont {Gustavsson},\
  and\ \citenamefont {Oliver}}]{Morten2020}%
  \BibitemOpen
  \bibfield  {author} {\bibinfo {author} {\bibfnamefont {M.}~\bibnamefont
  {Kjaergaard}}, \bibinfo {author} {\bibfnamefont {M.~E.}\ \bibnamefont
  {Schwartz}}, \bibinfo {author} {\bibfnamefont {J.}~\bibnamefont
  {Braumüller}}, \bibinfo {author} {\bibfnamefont {P.}~\bibnamefont {Krantz}},
  \bibinfo {author} {\bibfnamefont {J.~I.-J.}\ \bibnamefont {Wang}}, \bibinfo
  {author} {\bibfnamefont {S.}~\bibnamefont {Gustavsson}}, \ and\ \bibinfo
  {author} {\bibfnamefont {W.~D.}\ \bibnamefont {Oliver}},\ }\emph
  {Superconducting Qubits: Current State of Play},\ \href {\doibase
  10.1146/annurev-conmatphys-031119-050605} {\bibfield  {journal} {\bibinfo
  {journal} {Annu. Rev. Condens. Matter Phys}\ }\textbf {\bibinfo {volume}
  {11}},\ \bibinfo {pages} {369} (\bibinfo {year} {2020})}\BibitemShut
  {NoStop}%
\bibitem [{\citenamefont {Bloch}\ \emph {et~al.}(2008)\citenamefont {Bloch},
  \citenamefont {Dalibard},\ and\ \citenamefont {Zwerger}}]{Bloch08}%
  \BibitemOpen
  \bibfield  {author} {\bibinfo {author} {\bibfnamefont {I.}~\bibnamefont
  {Bloch}}, \bibinfo {author} {\bibfnamefont {J.}~\bibnamefont {Dalibard}}, \
  and\ \bibinfo {author} {\bibfnamefont {W.}~\bibnamefont {Zwerger}},\ }\emph
  {Many-body physics with ultracold gases},\ \href {\doibase
  10.1103/RevModPhys.80.885} {\bibfield  {journal} {\bibinfo  {journal} {Rev.
  Mod. Phys.}\ }\textbf {\bibinfo {volume} {80}},\ \bibinfo {pages} {885}
  (\bibinfo {year} {2008})}\BibitemShut {NoStop}%
\bibitem [{\citenamefont {Bloch}\ \emph {et~al.}(2012)\citenamefont {Bloch},
  \citenamefont {Dalibard},\ and\ \citenamefont {Nascimb{\`{e}}ne}}]{Bloch12}%
  \BibitemOpen
  \bibfield  {author} {\bibinfo {author} {\bibfnamefont {I.}~\bibnamefont
  {Bloch}}, \bibinfo {author} {\bibfnamefont {J.}~\bibnamefont {Dalibard}}, \
  and\ \bibinfo {author} {\bibfnamefont {S.}~\bibnamefont {Nascimb{\`{e}}ne}},\
  }\emph {Quantum simulations with ultracold quantum gases},\ \href {\doibase
  10.1038/nphys2259} {\bibfield  {journal} {\bibinfo  {journal} {Nat. Phys.}\
  }\textbf {\bibinfo {volume} {8}},\ \bibinfo {pages} {267} (\bibinfo {year}
  {2012})}\BibitemShut {NoStop}%
\bibitem [{\citenamefont {Gross}\ and\ \citenamefont {Bloch}(2017)}]{Gross17}%
  \BibitemOpen
  \bibfield  {author} {\bibinfo {author} {\bibfnamefont {C.}~\bibnamefont
  {Gross}}\ and\ \bibinfo {author} {\bibfnamefont {I.}~\bibnamefont {Bloch}},\
  }\emph {Quantum simulations with ultracold atoms in optical lattices},\ \href
  {\doibase 10.1126/science.aal3837} {\bibfield  {journal} {\bibinfo  {journal}
  {Science}\ }\textbf {\bibinfo {volume} {357}},\ \bibinfo {pages} {995}
  (\bibinfo {year} {2017})}\BibitemShut {NoStop}%
\bibitem [{\citenamefont {Browaeys}\ and\ \citenamefont
  {Lahaye}(2020)}]{Browaeys2020}%
  \BibitemOpen
  \bibfield  {author} {\bibinfo {author} {\bibfnamefont {A.}~\bibnamefont
  {Browaeys}}\ and\ \bibinfo {author} {\bibfnamefont {T.}~\bibnamefont
  {Lahaye}},\ }\emph {{Many-body physics with individually controlled Rydberg
  atoms.}},\ \href {\doibase 10.1038/s41567-019-0733-z} {\bibfield  {journal}
  {\bibinfo  {journal} {Nat. Phys.}\ }\textbf {\bibinfo {volume} {16}},\
  \bibinfo {pages} {132} (\bibinfo {year} {2020})}\BibitemShut {NoStop}%
\bibitem [{\citenamefont {Yan}\ \emph {et~al.}(2013)\citenamefont {Yan},
  \citenamefont {Moses}, \citenamefont {Gadway}, \citenamefont {Covey},
  \citenamefont {Hazzard}, \citenamefont {Rey}, \citenamefont {Jin},\ and\
  \citenamefont {Ye}}]{Yan2013}%
  \BibitemOpen
  \bibfield  {author} {\bibinfo {author} {\bibfnamefont {B.}~\bibnamefont
  {Yan}}, \bibinfo {author} {\bibfnamefont {S.~A.}\ \bibnamefont {Moses}},
  \bibinfo {author} {\bibfnamefont {B.}~\bibnamefont {Gadway}}, \bibinfo
  {author} {\bibfnamefont {J.~P.}\ \bibnamefont {Covey}}, \bibinfo {author}
  {\bibfnamefont {K.~R.~A.}\ \bibnamefont {Hazzard}}, \bibinfo {author}
  {\bibfnamefont {A.~M.}\ \bibnamefont {Rey}}, \bibinfo {author} {\bibfnamefont
  {D.~S.}\ \bibnamefont {Jin}}, \ and\ \bibinfo {author} {\bibfnamefont
  {J.}~\bibnamefont {Ye}},\ }\emph {Observation of dipolar spin-exchange
  interactions with lattice-confined polar molecules},\ \href {\doibase
  10.1038/nature12483} {\bibfield  {journal} {\bibinfo  {journal} {Nature}\
  }\textbf {\bibinfo {volume} {501}},\ \bibinfo {pages} {521} (\bibinfo {year}
  {2013})}\BibitemShut {NoStop}%
\bibitem [{\citenamefont {Zhou}\ \emph {et~al.}(2011)\citenamefont {Zhou},
  \citenamefont {Ortner},\ and\ \citenamefont {Rabl}}]{Zhou2011}%
  \BibitemOpen
  \bibfield  {author} {\bibinfo {author} {\bibfnamefont {Y.~L.}\ \bibnamefont
  {Zhou}}, \bibinfo {author} {\bibfnamefont {M.}~\bibnamefont {Ortner}}, \ and\
  \bibinfo {author} {\bibfnamefont {P.}~\bibnamefont {Rabl}},\ }\emph
  {Long-range and frustrated spin-spin interactions in crystals of cold polar
  molecules},\ \href {\doibase 10.1103/PhysRevA.84.052332} {\bibfield
  {journal} {\bibinfo  {journal} {Phys. Rev. A}\ }\textbf {\bibinfo {volume}
  {84}},\ \bibinfo {pages} {052332} (\bibinfo {year} {2011})}\BibitemShut
  {NoStop}%
\bibitem [{\citenamefont {Bernien}\ \emph {et~al.}(2017)\citenamefont
  {Bernien}, \citenamefont {Schwartz}, \citenamefont {Keesling}, \citenamefont
  {Levine}, \citenamefont {Omran}, \citenamefont {Pichler}, \citenamefont
  {Choi}, \citenamefont {Zibrov}, \citenamefont {Endres}, \citenamefont
  {Greiner}, \citenamefont {Vuletić},\ and\ \citenamefont
  {Lukin}}]{Bernien2017}%
  \BibitemOpen
  \bibfield  {author} {\bibinfo {author} {\bibfnamefont {H.}~\bibnamefont
  {Bernien}}, \bibinfo {author} {\bibfnamefont {S.}~\bibnamefont {Schwartz}},
  \bibinfo {author} {\bibfnamefont {A.}~\bibnamefont {Keesling}}, \bibinfo
  {author} {\bibfnamefont {H.}~\bibnamefont {Levine}}, \bibinfo {author}
  {\bibfnamefont {A.}~\bibnamefont {Omran}}, \bibinfo {author} {\bibfnamefont
  {H.}~\bibnamefont {Pichler}}, \bibinfo {author} {\bibfnamefont
  {S.}~\bibnamefont {Choi}}, \bibinfo {author} {\bibfnamefont {A.~S.}\
  \bibnamefont {Zibrov}}, \bibinfo {author} {\bibfnamefont {M.}~\bibnamefont
  {Endres}}, \bibinfo {author} {\bibfnamefont {M.}~\bibnamefont {Greiner}},
  \bibinfo {author} {\bibfnamefont {V.}~\bibnamefont {Vuletić}}, \ and\
  \bibinfo {author} {\bibfnamefont {M.~D.}\ \bibnamefont {Lukin}},\ }\emph
  {{Probing many-body dynamics on a 51-atom quantum simulator}},\ \href
  {\doibase 10.1038/nature24622} {\bibfield  {journal} {\bibinfo  {journal}
  {Nature}\ }\textbf {\bibinfo {volume} {551}},\ \bibinfo {pages} {579}
  (\bibinfo {year} {2017})}\BibitemShut {NoStop}%
\bibitem [{\citenamefont {Keesling}\ \emph {et~al.}(2019)\citenamefont
  {Keesling}, \citenamefont {Omran}, \citenamefont {Levine}, \citenamefont
  {Bernien}, \citenamefont {Pichler}, \citenamefont {Choi}, \citenamefont
  {Samajdar}, \citenamefont {Schwartz}, \citenamefont {Silvi}, \citenamefont
  {Sachdev}, \citenamefont {Zoller}, \citenamefont {Endres}, \citenamefont
  {Greiner}, \citenamefont {Vuleti{\'{c}}},\ and\ \citenamefont
  {Lukin}}]{Keesling2019}%
  \BibitemOpen
  \bibfield  {author} {\bibinfo {author} {\bibfnamefont {A.}~\bibnamefont
  {Keesling}}, \bibinfo {author} {\bibfnamefont {A.}~\bibnamefont {Omran}},
  \bibinfo {author} {\bibfnamefont {H.}~\bibnamefont {Levine}}, \bibinfo
  {author} {\bibfnamefont {H.}~\bibnamefont {Bernien}}, \bibinfo {author}
  {\bibfnamefont {H.}~\bibnamefont {Pichler}}, \bibinfo {author} {\bibfnamefont
  {S.}~\bibnamefont {Choi}}, \bibinfo {author} {\bibfnamefont {R.}~\bibnamefont
  {Samajdar}}, \bibinfo {author} {\bibfnamefont {S.}~\bibnamefont {Schwartz}},
  \bibinfo {author} {\bibfnamefont {P.}~\bibnamefont {Silvi}}, \bibinfo
  {author} {\bibfnamefont {S.}~\bibnamefont {Sachdev}}, \bibinfo {author}
  {\bibfnamefont {P.}~\bibnamefont {Zoller}}, \bibinfo {author} {\bibfnamefont
  {M.}~\bibnamefont {Endres}}, \bibinfo {author} {\bibfnamefont
  {M.}~\bibnamefont {Greiner}}, \bibinfo {author} {\bibfnamefont
  {V.}~\bibnamefont {Vuleti{\'{c}}}}, \ and\ \bibinfo {author} {\bibfnamefont
  {M.~D.}\ \bibnamefont {Lukin}},\ }\emph {{Quantum Kibble--Zurek mechanism and
  critical dynamics on a programmable Rydberg simulator}},\ \href {\doibase
  10.1038/s41586-019-1070-1} {\bibfield  {journal} {\bibinfo  {journal}
  {Nature}\ }\textbf {\bibinfo {volume} {568}},\ \bibinfo {pages} {207}
  (\bibinfo {year} {2019})}\BibitemShut {NoStop}%
\bibitem [{\citenamefont {Lienhard}\ \emph {et~al.}(2018)\citenamefont
  {Lienhard}, \citenamefont {de~L\'es\'eleuc}, \citenamefont {Barredo},
  \citenamefont {Lahaye}, \citenamefont {Browaeys}, \citenamefont {Schuler},
  \citenamefont {Henry},\ and\ \citenamefont {L\"auchli}}]{Lienhard2018}%
  \BibitemOpen
  \bibfield  {author} {\bibinfo {author} {\bibfnamefont {V.}~\bibnamefont
  {Lienhard}}, \bibinfo {author} {\bibfnamefont {S.}~\bibnamefont
  {de~L\'es\'eleuc}}, \bibinfo {author} {\bibfnamefont {D.}~\bibnamefont
  {Barredo}}, \bibinfo {author} {\bibfnamefont {T.}~\bibnamefont {Lahaye}},
  \bibinfo {author} {\bibfnamefont {A.}~\bibnamefont {Browaeys}}, \bibinfo
  {author} {\bibfnamefont {M.}~\bibnamefont {Schuler}}, \bibinfo {author}
  {\bibfnamefont {L.-P.}\ \bibnamefont {Henry}}, \ and\ \bibinfo {author}
  {\bibfnamefont {A.~M.}\ \bibnamefont {L\"auchli}},\ }\emph {{Observing the
  Space- and Time-Dependent Growth of Correlations in Dynamically Tuned
  Synthetic Ising Antiferromagnets}},\ \href {\doibase
  10.1103/PhysRevX.8.021070} {\bibfield  {journal} {\bibinfo  {journal} {Phys.
  Rev. X}\ }\textbf {\bibinfo {volume} {8}},\ \bibinfo {pages} {021070}
  (\bibinfo {year} {2018})}\BibitemShut {NoStop}%
\bibitem [{\citenamefont {Levine}\ \emph {et~al.}(2019)\citenamefont {Levine},
  \citenamefont {Keesling}, \citenamefont {Semeghini}, \citenamefont {Omran},
  \citenamefont {Wang}, \citenamefont {Ebadi}, \citenamefont {Bernien},
  \citenamefont {Greiner}, \citenamefont {Vuleti\ifmmode~\acute{c}\else
  \'{c}\fi{}}, \citenamefont {Pichler},\ and\ \citenamefont
  {Lukin}}]{Levine2019}%
  \BibitemOpen
  \bibfield  {author} {\bibinfo {author} {\bibfnamefont {H.}~\bibnamefont
  {Levine}}, \bibinfo {author} {\bibfnamefont {A.}~\bibnamefont {Keesling}},
  \bibinfo {author} {\bibfnamefont {G.}~\bibnamefont {Semeghini}}, \bibinfo
  {author} {\bibfnamefont {A.}~\bibnamefont {Omran}}, \bibinfo {author}
  {\bibfnamefont {T.~T.}\ \bibnamefont {Wang}}, \bibinfo {author}
  {\bibfnamefont {S.}~\bibnamefont {Ebadi}}, \bibinfo {author} {\bibfnamefont
  {H.}~\bibnamefont {Bernien}}, \bibinfo {author} {\bibfnamefont
  {M.}~\bibnamefont {Greiner}}, \bibinfo {author} {\bibfnamefont
  {V.}~\bibnamefont {Vuleti\ifmmode~\acute{c}\else \'{c}\fi{}}}, \bibinfo
  {author} {\bibfnamefont {H.}~\bibnamefont {Pichler}}, \ and\ \bibinfo
  {author} {\bibfnamefont {M.~D.}\ \bibnamefont {Lukin}},\ }\emph {Parallel
  Implementation of High-Fidelity Multiqubit Gates with Neutral Atoms},\ \href
  {\doibase 10.1103/PhysRevLett.123.170503} {\bibfield  {journal} {\bibinfo
  {journal} {Phys. Rev. Lett.}\ }\textbf {\bibinfo {volume} {123}},\ \bibinfo
  {pages} {170503} (\bibinfo {year} {2019})}\BibitemShut {NoStop}%
\bibitem [{\citenamefont {Omran}\ \emph {et~al.}(2019)\citenamefont {Omran},
  \citenamefont {Levine}, \citenamefont {Keesling}, \citenamefont {Semeghini},
  \citenamefont {Wang}, \citenamefont {Ebadi}, \citenamefont {Bernien},
  \citenamefont {Zibrov}, \citenamefont {Pichler}, \citenamefont {Choi},
  \citenamefont {Cui}, \citenamefont {Rossignolo}, \citenamefont {Rembold},
  \citenamefont {Montangero}, \citenamefont {Calarco}, \citenamefont {Endres},
  \citenamefont {Greiner}, \citenamefont {Vuleti{\'c}},\ and\ \citenamefont
  {Lukin}}]{Omran2019}%
  \BibitemOpen
  \bibfield  {author} {\bibinfo {author} {\bibfnamefont {A.}~\bibnamefont
  {Omran}}, \bibinfo {author} {\bibfnamefont {H.}~\bibnamefont {Levine}},
  \bibinfo {author} {\bibfnamefont {A.}~\bibnamefont {Keesling}}, \bibinfo
  {author} {\bibfnamefont {G.}~\bibnamefont {Semeghini}}, \bibinfo {author}
  {\bibfnamefont {T.~T.}\ \bibnamefont {Wang}}, \bibinfo {author}
  {\bibfnamefont {S.}~\bibnamefont {Ebadi}}, \bibinfo {author} {\bibfnamefont
  {H.}~\bibnamefont {Bernien}}, \bibinfo {author} {\bibfnamefont {A.~S.}\
  \bibnamefont {Zibrov}}, \bibinfo {author} {\bibfnamefont {H.}~\bibnamefont
  {Pichler}}, \bibinfo {author} {\bibfnamefont {S.}~\bibnamefont {Choi}},
  \bibinfo {author} {\bibfnamefont {J.}~\bibnamefont {Cui}}, \bibinfo {author}
  {\bibfnamefont {M.}~\bibnamefont {Rossignolo}}, \bibinfo {author}
  {\bibfnamefont {P.}~\bibnamefont {Rembold}}, \bibinfo {author} {\bibfnamefont
  {S.}~\bibnamefont {Montangero}}, \bibinfo {author} {\bibfnamefont
  {T.}~\bibnamefont {Calarco}}, \bibinfo {author} {\bibfnamefont
  {M.}~\bibnamefont {Endres}}, \bibinfo {author} {\bibfnamefont
  {M.}~\bibnamefont {Greiner}}, \bibinfo {author} {\bibfnamefont
  {V.}~\bibnamefont {Vuleti{\'c}}}, \ and\ \bibinfo {author} {\bibfnamefont
  {M.~D.}\ \bibnamefont {Lukin}},\ }\emph {Generation and manipulation of
  Schr{\"o}dinger cat states in Rydberg atom arrays},\ \href {\doibase
  10.1126/science.aax9743} {\bibfield  {journal} {\bibinfo  {journal}
  {Science}\ }\textbf {\bibinfo {volume} {365}},\ \bibinfo {pages} {570}
  (\bibinfo {year} {2019})}\BibitemShut {NoStop}%
\bibitem [{\citenamefont {Madjarov}\ \emph {et~al.}(2020)\citenamefont
  {Madjarov}, \citenamefont {Covey}, \citenamefont {Shaw}, \citenamefont
  {Choi}, \citenamefont {Kale}, \citenamefont {Cooper}, \citenamefont
  {Pichler}, \citenamefont {Schkolnik}, \citenamefont {Williams},\ and\
  \citenamefont {Endres}}]{Madjarov2020}%
  \BibitemOpen
  \bibfield  {author} {\bibinfo {author} {\bibfnamefont {I.~S.}\ \bibnamefont
  {Madjarov}}, \bibinfo {author} {\bibfnamefont {J.~P.}\ \bibnamefont {Covey}},
  \bibinfo {author} {\bibfnamefont {A.~L.}\ \bibnamefont {Shaw}}, \bibinfo
  {author} {\bibfnamefont {J.}~\bibnamefont {Choi}}, \bibinfo {author}
  {\bibfnamefont {A.}~\bibnamefont {Kale}}, \bibinfo {author} {\bibfnamefont
  {A.}~\bibnamefont {Cooper}}, \bibinfo {author} {\bibfnamefont
  {H.}~\bibnamefont {Pichler}}, \bibinfo {author} {\bibfnamefont
  {V.}~\bibnamefont {Schkolnik}}, \bibinfo {author} {\bibfnamefont {J.~R.}\
  \bibnamefont {Williams}}, \ and\ \bibinfo {author} {\bibfnamefont
  {M.}~\bibnamefont {Endres}},\ }\emph {High-fidelity entanglement and
  detection of alkaline-earth Rydberg atoms},\ \href {\doibase
  10.1038/s41567-020-0903-z} {\bibfield  {journal} {\bibinfo  {journal} {Nat.
  Phys.}\ }\textbf {\bibinfo {volume} {16}},\ \bibinfo {pages} {857} (\bibinfo
  {year} {2020})}\BibitemShut {NoStop}%
\bibitem [{\citenamefont {Schau{\ss}}\ \emph {et~al.}(2015)\citenamefont
  {Schau{\ss}}, \citenamefont {Zeiher}, \citenamefont {Fukuhara}, \citenamefont
  {Hild}, \citenamefont {Cheneau}, \citenamefont {Macr{\`\i}}, \citenamefont
  {Pohl}, \citenamefont {Bloch},\ and\ \citenamefont {Gross}}]{Schauss2015}%
  \BibitemOpen
  \bibfield  {author} {\bibinfo {author} {\bibfnamefont {P.}~\bibnamefont
  {Schau{\ss}}}, \bibinfo {author} {\bibfnamefont {J.}~\bibnamefont {Zeiher}},
  \bibinfo {author} {\bibfnamefont {T.}~\bibnamefont {Fukuhara}}, \bibinfo
  {author} {\bibfnamefont {S.}~\bibnamefont {Hild}}, \bibinfo {author}
  {\bibfnamefont {M.}~\bibnamefont {Cheneau}}, \bibinfo {author} {\bibfnamefont
  {T.}~\bibnamefont {Macr{\`\i}}}, \bibinfo {author} {\bibfnamefont
  {T.}~\bibnamefont {Pohl}}, \bibinfo {author} {\bibfnamefont {I.}~\bibnamefont
  {Bloch}}, \ and\ \bibinfo {author} {\bibfnamefont {C.}~\bibnamefont
  {Gross}},\ }\emph {Crystallization in Ising quantum magnets},\ \href
  {\doibase 10.1126/science.1258351} {\bibfield  {journal} {\bibinfo  {journal}
  {Science}\ }\textbf {\bibinfo {volume} {347}},\ \bibinfo {pages} {1455}
  (\bibinfo {year} {2015})}\BibitemShut {NoStop}%
\bibitem [{\citenamefont {Guardado-Sanchez}\ \emph {et~al.}(2018)\citenamefont
  {Guardado-Sanchez}, \citenamefont {Brown}, \citenamefont {Mitra},
  \citenamefont {Devakul}, \citenamefont {Huse}, \citenamefont {Schau\ss{}},\
  and\ \citenamefont {Bakr}}]{GuardadoSanchez2018}%
  \BibitemOpen
  \bibfield  {author} {\bibinfo {author} {\bibfnamefont {E.}~\bibnamefont
  {Guardado-Sanchez}}, \bibinfo {author} {\bibfnamefont {P.~T.}\ \bibnamefont
  {Brown}}, \bibinfo {author} {\bibfnamefont {D.}~\bibnamefont {Mitra}},
  \bibinfo {author} {\bibfnamefont {T.}~\bibnamefont {Devakul}}, \bibinfo
  {author} {\bibfnamefont {D.~A.}\ \bibnamefont {Huse}}, \bibinfo {author}
  {\bibfnamefont {P.}~\bibnamefont {Schau\ss{}}}, \ and\ \bibinfo {author}
  {\bibfnamefont {W.~S.}\ \bibnamefont {Bakr}},\ }\emph {Probing the Quench
  Dynamics of Antiferromagnetic Correlations in a 2D Quantum Ising Spin
  System},\ \href {\doibase 10.1103/PhysRevX.8.021069} {\bibfield  {journal}
  {\bibinfo  {journal} {Phys. Rev. X}\ }\textbf {\bibinfo {volume} {8}},\
  \bibinfo {pages} {021069} (\bibinfo {year} {2018})}\BibitemShut {NoStop}%
\bibitem [{\citenamefont {Song}\ \emph {et~al.}(2020)\citenamefont {Song},
  \citenamefont {Kim}, \citenamefont {Hwang}, \citenamefont {Lee},\ and\
  \citenamefont {Ahn}}]{Song2020}%
  \BibitemOpen
  \bibfield  {author} {\bibinfo {author} {\bibfnamefont {Y.}~\bibnamefont
  {Song}}, \bibinfo {author} {\bibfnamefont {M.}~\bibnamefont {Kim}}, \bibinfo
  {author} {\bibfnamefont {H.}~\bibnamefont {Hwang}}, \bibinfo {author}
  {\bibfnamefont {W.}~\bibnamefont {Lee}}, \ and\ \bibinfo {author}
  {\bibfnamefont {J.}~\bibnamefont {Ahn}},\ }\emph {Quantum annealing of
  Cayley-tree Ising spins at small scales},\ \href
  {https://arxiv.org/abs/2011.01653} {\bibfield  {journal} {\bibinfo  {journal}
  {arXiv:2011.01653}\ } (\bibinfo {year} {2020})}\BibitemShut {NoStop}%
\bibitem [{\citenamefont {Wannier}(1950)}]{Wannier1950}%
  \BibitemOpen
  \bibfield  {author} {\bibinfo {author} {\bibfnamefont {G.~H.}\ \bibnamefont
  {Wannier}},\ }\emph {Antiferromagnetism. The Triangular Ising Net},\ \href
  {\doibase 10.1103/PhysRev.79.357} {\bibfield  {journal} {\bibinfo  {journal}
  {Phys. Rev.}\ }\textbf {\bibinfo {volume} {79}},\ \bibinfo {pages} {357}
  (\bibinfo {year} {1950})}\BibitemShut {NoStop}%
\bibitem [{\citenamefont {Jaksch}\ \emph {et~al.}(2000)\citenamefont {Jaksch},
  \citenamefont {Cirac}, \citenamefont {Zoller}, \citenamefont {Rolston},
  \citenamefont {C\^ot\'e},\ and\ \citenamefont {Lukin}}]{Jaksch2000}%
  \BibitemOpen
  \bibfield  {author} {\bibinfo {author} {\bibfnamefont {D.}~\bibnamefont
  {Jaksch}}, \bibinfo {author} {\bibfnamefont {J.~I.}\ \bibnamefont {Cirac}},
  \bibinfo {author} {\bibfnamefont {P.}~\bibnamefont {Zoller}}, \bibinfo
  {author} {\bibfnamefont {S.~L.}\ \bibnamefont {Rolston}}, \bibinfo {author}
  {\bibfnamefont {R.}~\bibnamefont {C\^ot\'e}}, \ and\ \bibinfo {author}
  {\bibfnamefont {M.~D.}\ \bibnamefont {Lukin}},\ }\emph {Fast Quantum Gates
  for Neutral Atoms},\ \href {\doibase 10.1103/PhysRevLett.85.2208} {\bibfield
  {journal} {\bibinfo  {journal} {Phys. Rev. Lett.}\ }\textbf {\bibinfo
  {volume} {85}},\ \bibinfo {pages} {2208} (\bibinfo {year}
  {2000})}\BibitemShut {NoStop}%
\bibitem [{\citenamefont {Schymik}\ \emph {et~al.}(2020)\citenamefont
  {Schymik}, \citenamefont {Lienhard}, \citenamefont {Barredo}, \citenamefont
  {Scholl}, \citenamefont {Williams}, \citenamefont {Browaeys},\ and\
  \citenamefont {Lahaye}}]{Schymik2020}%
  \BibitemOpen
  \bibfield  {author} {\bibinfo {author} {\bibfnamefont {K.-N.}\ \bibnamefont
  {Schymik}}, \bibinfo {author} {\bibfnamefont {V.}~\bibnamefont {Lienhard}},
  \bibinfo {author} {\bibfnamefont {D.}~\bibnamefont {Barredo}}, \bibinfo
  {author} {\bibfnamefont {P.}~\bibnamefont {Scholl}}, \bibinfo {author}
  {\bibfnamefont {H.}~\bibnamefont {Williams}}, \bibinfo {author}
  {\bibfnamefont {A.}~\bibnamefont {Browaeys}}, \ and\ \bibinfo {author}
  {\bibfnamefont {T.}~\bibnamefont {Lahaye}},\ }\emph {Enhanced atom-by-atom
  assembly of arbitrary tweezer arrays},\ \href {\doibase
  10.1103/PhysRevA.102.063107} {\bibfield  {journal} {\bibinfo  {journal}
  {Phys. Rev. A}\ }\textbf {\bibinfo {volume} {102}},\ \bibinfo {pages}
  {063107} (\bibinfo {year} {2020})}\BibitemShut {NoStop}%
\bibitem [{\citenamefont {Levine}\ \emph {et~al.}(2018)\citenamefont {Levine},
  \citenamefont {Keesling}, \citenamefont {Omran}, \citenamefont {Bernien},
  \citenamefont {Schwartz}, \citenamefont {Zibrov}, \citenamefont {Endres},
  \citenamefont {Greiner}, \citenamefont {Vuleti\ifmmode~\acute{c}\else
  \'{c}\fi{}},\ and\ \citenamefont {Lukin}}]{Levine2018}%
  \BibitemOpen
  \bibfield  {author} {\bibinfo {author} {\bibfnamefont {H.}~\bibnamefont
  {Levine}}, \bibinfo {author} {\bibfnamefont {A.}~\bibnamefont {Keesling}},
  \bibinfo {author} {\bibfnamefont {A.}~\bibnamefont {Omran}}, \bibinfo
  {author} {\bibfnamefont {H.}~\bibnamefont {Bernien}}, \bibinfo {author}
  {\bibfnamefont {S.}~\bibnamefont {Schwartz}}, \bibinfo {author}
  {\bibfnamefont {A.~S.}\ \bibnamefont {Zibrov}}, \bibinfo {author}
  {\bibfnamefont {M.}~\bibnamefont {Endres}}, \bibinfo {author} {\bibfnamefont
  {M.}~\bibnamefont {Greiner}}, \bibinfo {author} {\bibfnamefont
  {V.}~\bibnamefont {Vuleti\ifmmode~\acute{c}\else \'{c}\fi{}}}, \ and\
  \bibinfo {author} {\bibfnamefont {M.~D.}\ \bibnamefont {Lukin}},\ }\emph
  {High-Fidelity Control and Entanglement of Rydberg-Atom Qubits},\ \href
  {\doibase 10.1103/PhysRevLett.121.123603} {\bibfield  {journal} {\bibinfo
  {journal} {Phys. Rev. Lett.}\ }\textbf {\bibinfo {volume} {121}},\ \bibinfo
  {pages} {123603} (\bibinfo {year} {2018})}\BibitemShut {NoStop}%
\bibitem [{\citenamefont {Janke}\ and\ \citenamefont
  {Villanova}(1997)}]{Janke1997}%
  \BibitemOpen
  \bibfield  {author} {\bibinfo {author} {\bibfnamefont {W.}~\bibnamefont
  {Janke}}\ and\ \bibinfo {author} {\bibfnamefont {R.}~\bibnamefont
  {Villanova}},\ }\emph {Three-dimensional 3-state Potts model revisited with
  new techniques},\ \href {\doibase
  https://doi.org/10.1016/S0550-3213(96)00710-9} {\bibfield  {journal}
  {\bibinfo  {journal} {Nucl. Phys. B}\ }\textbf {\bibinfo {volume} {489}},\
  \bibinfo {pages} {679 } (\bibinfo {year} {1997})}\BibitemShut {NoStop}%
\bibitem [{\citenamefont {Laumann}\ \emph {et~al.}(2012)\citenamefont
  {Laumann}, \citenamefont {Moessner}, \citenamefont {Scardicchio},\ and\
  \citenamefont {Sondhi}}]{Laumann2012}%
  \BibitemOpen
  \bibfield  {author} {\bibinfo {author} {\bibfnamefont {C.~R.}\ \bibnamefont
  {Laumann}}, \bibinfo {author} {\bibfnamefont {R.}~\bibnamefont {Moessner}},
  \bibinfo {author} {\bibfnamefont {A.}~\bibnamefont {Scardicchio}}, \ and\
  \bibinfo {author} {\bibfnamefont {S.~L.}\ \bibnamefont {Sondhi}},\ }\emph
  {Quantum Adiabatic Algorithm and Scaling of Gaps at First-Order Quantum Phase
  Transitions},\ \href {\doibase 10.1103/PhysRevLett.109.030502} {\bibfield
  {journal} {\bibinfo  {journal} {Phys. Rev. Lett.}\ }\textbf {\bibinfo
  {volume} {109}},\ \bibinfo {pages} {030502} (\bibinfo {year}
  {2012})}\BibitemShut {NoStop}%
\bibitem [{\citenamefont {Fey}\ \emph {et~al.}(2019)\citenamefont {Fey},
  \citenamefont {Kapfer},\ and\ \citenamefont {Schmidt}}]{Fey2019}%
  \BibitemOpen
  \bibfield  {author} {\bibinfo {author} {\bibfnamefont {S.}~\bibnamefont
  {Fey}}, \bibinfo {author} {\bibfnamefont {S.~C.}\ \bibnamefont {Kapfer}}, \
  and\ \bibinfo {author} {\bibfnamefont {K.~P.}\ \bibnamefont {Schmidt}},\
  }\emph {Quantum Criticality of Two-Dimensional Quantum Magnets with
  Long-Range Interactions},\ \href {\doibase 10.1103/PhysRevLett.122.017203}
  {\bibfield  {journal} {\bibinfo  {journal} {Phys. Rev. Lett.}\ }\textbf
  {\bibinfo {volume} {122}},\ \bibinfo {pages} {017203} (\bibinfo {year}
  {2019})}\BibitemShut {NoStop}%
\bibitem [{\citenamefont {Samajdar}\ \emph {et~al.}(2020)\citenamefont
  {Samajdar}, \citenamefont {Ho}, \citenamefont {Pichler}, \citenamefont
  {Lukin},\ and\ \citenamefont {Sachdev}}]{Samajdar2020}%
  \BibitemOpen
  \bibfield  {author} {\bibinfo {author} {\bibfnamefont {R.}~\bibnamefont
  {Samajdar}}, \bibinfo {author} {\bibfnamefont {W.~W.}\ \bibnamefont {Ho}},
  \bibinfo {author} {\bibfnamefont {H.}~\bibnamefont {Pichler}}, \bibinfo
  {author} {\bibfnamefont {M.~D.}\ \bibnamefont {Lukin}}, \ and\ \bibinfo
  {author} {\bibfnamefont {S.}~\bibnamefont {Sachdev}},\ }\emph {Complex
  Density Wave Orders and Quantum Phase Transitions in a Model of
  Square-Lattice Rydberg Atom Arrays},\ \href {\doibase
  10.1103/PhysRevLett.124.103601} {\bibfield  {journal} {\bibinfo  {journal}
  {Phys. Rev. Lett.}\ }\textbf {\bibinfo {volume} {124}},\ \bibinfo {pages}
  {103601} (\bibinfo {year} {2020})}\BibitemShut {NoStop}%
\bibitem [{Note1()}]{Note1}%
  \BibitemOpen
  \bibinfo {note} {The residual anisotropy observed is due to the finite size
  of the excitation beams, which is comparable to the width of the array
  ($130~\mu $m).}\BibitemShut {Stop}%
\bibitem [{\citenamefont {Stoli}\ and\ \citenamefont {Domb}(1979)}]{Stoli1979}%
  \BibitemOpen
  \bibfield  {author} {\bibinfo {author} {\bibfnamefont {E.}~\bibnamefont
  {Stoli}}\ and\ \bibinfo {author} {\bibfnamefont {C.}~\bibnamefont {Domb}},\
  }\emph {Shape and size of two-dimensional percolation clusters with and
  without correlations},\ \href {\doibase 10.1088/0305-4470/12/10/029}
  {\bibfield  {journal} {\bibinfo  {journal} {J. Phys. A}\ }\textbf {\bibinfo
  {volume} {12}},\ \bibinfo {pages} {1843} (\bibinfo {year}
  {1979})}\BibitemShut {NoStop}%
\bibitem [{\citenamefont {Kim}\ \emph {et~al.}(2018)\citenamefont {Kim},
  \citenamefont {Park}, \citenamefont {Kim}, \citenamefont {Sim},\ and\
  \citenamefont {Ahn}}]{Kim2018}%
  \BibitemOpen
  \bibfield  {author} {\bibinfo {author} {\bibfnamefont {H.}~\bibnamefont
  {Kim}}, \bibinfo {author} {\bibfnamefont {Y.}~\bibnamefont {Park}}, \bibinfo
  {author} {\bibfnamefont {K.}~\bibnamefont {Kim}}, \bibinfo {author}
  {\bibfnamefont {H.-S.}\ \bibnamefont {Sim}}, \ and\ \bibinfo {author}
  {\bibfnamefont {J.}~\bibnamefont {Ahn}},\ }\emph {Detailed Balance of
  Thermalization Dynamics in Rydberg-Atom Quantum Simulators},\ \href {\doibase
  10.1103/PhysRevLett.120.180502} {\bibfield  {journal} {\bibinfo  {journal}
  {Phys. Rev. Lett.}\ }\textbf {\bibinfo {volume} {120}},\ \bibinfo {pages}
  {180502} (\bibinfo {year} {2018})}\BibitemShut {NoStop}%
\bibitem [{Note2()}]{Note2}%
  \BibitemOpen
  \bibinfo {note} {One could also use other observables to match a hypothetical
  temperature, but since temperature is the variable conjugate to the energy in
  thermodynamics, it is the most natural choice.}\BibitemShut {Stop}%
\bibitem [{\citenamefont {Pichler}\ \emph {et~al.}(2018)\citenamefont
  {Pichler}, \citenamefont {Wang}, \citenamefont {Zhou}, \citenamefont {Choi},\
  and\ \citenamefont {Lukin}}]{Pichler2018}%
  \BibitemOpen
  \bibfield  {author} {\bibinfo {author} {\bibfnamefont {H.}~\bibnamefont
  {Pichler}}, \bibinfo {author} {\bibfnamefont {S.-T.}\ \bibnamefont {Wang}},
  \bibinfo {author} {\bibfnamefont {L.}~\bibnamefont {Zhou}}, \bibinfo {author}
  {\bibfnamefont {S.}~\bibnamefont {Choi}}, \ and\ \bibinfo {author}
  {\bibfnamefont {M.~D.}\ \bibnamefont {Lukin}},\ }\emph {Quantum Optimization
  for Maximum Independent Set Using Rydberg Atom Arrays},\ \href
  {https://arxiv.org/abs/1808.10816} {\bibfield  {journal} {\bibinfo  {journal}
  {arXiv:1808.10816}\ } (\bibinfo {year} {2018})}\BibitemShut {NoStop}%
\bibitem [{\citenamefont {Henriet}(2020)}]{Henriet2020a}%
  \BibitemOpen
  \bibfield  {author} {\bibinfo {author} {\bibfnamefont {L.}~\bibnamefont
  {Henriet}},\ }\emph {Robustness to spontaneous emission of a variational
  quantum algorithm},\ \href {\doibase 10.1103/PhysRevA.101.012335} {\bibfield
  {journal} {\bibinfo  {journal} {Phys. Rev. A}\ }\textbf {\bibinfo {volume}
  {101}},\ \bibinfo {pages} {012335} (\bibinfo {year} {2020})}\BibitemShut
  {NoStop}%
\bibitem [{\citenamefont {Serret}\ \emph {et~al.}(2020)\citenamefont {Serret},
  \citenamefont {Marchand},\ and\ \citenamefont {Ayral}}]{Serret2020}%
  \BibitemOpen
  \bibfield  {author} {\bibinfo {author} {\bibfnamefont {M.~F.}\ \bibnamefont
  {Serret}}, \bibinfo {author} {\bibfnamefont {B.}~\bibnamefont {Marchand}}, \
  and\ \bibinfo {author} {\bibfnamefont {T.}~\bibnamefont {Ayral}},\ }\emph
  {Solving optimization problems with Rydberg analog quantum computers:
  Realistic requirements for quantum advantage using noisy simulation and
  classical benchmarks},\ \href {\doibase 10.1103/PhysRevA.102.052617}
  {\bibfield  {journal} {\bibinfo  {journal} {Phys. Rev. A}\ }\textbf {\bibinfo
  {volume} {102}},\ \bibinfo {pages} {052617} (\bibinfo {year}
  {2020})}\BibitemShut {NoStop}%
\bibitem [{\citenamefont {Villain}\ \emph {et~al.}(1980)\citenamefont
  {Villain}, \citenamefont {Bidaux}, \citenamefont {Carton},\ and\
  \citenamefont {Conte}}]{Villain1980}%
  \BibitemOpen
  \bibfield  {author} {\bibinfo {author} {\bibfnamefont {J.}~\bibnamefont
  {Villain}}, \bibinfo {author} {\bibfnamefont {R.}~\bibnamefont {Bidaux}},
  \bibinfo {author} {\bibfnamefont {J.-P.}\ \bibnamefont {Carton}}, \ and\
  \bibinfo {author} {\bibfnamefont {R.}~\bibnamefont {Conte}},\ }\emph {Order
  as an effect of disorder},\ \href {\doibase
  10.1051/jphys:0198000410110126300} {\bibfield  {journal} {\bibinfo  {journal}
  {J. Phys. France}\ }\textbf {\bibinfo {volume} {41}},\ \bibinfo {pages}
  {1263} (\bibinfo {year} {1980})}\BibitemShut {NoStop}%
\bibitem [{\citenamefont {Moessner}\ \emph {et~al.}(2000)\citenamefont
  {Moessner}, \citenamefont {Sondhi},\ and\ \citenamefont
  {Chandra}}]{Moessner2000}%
  \BibitemOpen
  \bibfield  {author} {\bibinfo {author} {\bibfnamefont {R.}~\bibnamefont
  {Moessner}}, \bibinfo {author} {\bibfnamefont {S.~L.}\ \bibnamefont
  {Sondhi}}, \ and\ \bibinfo {author} {\bibfnamefont {P.}~\bibnamefont
  {Chandra}},\ }\emph {{Two-Dimensional Periodic Frustrated Ising Models in a
  Transverse Field}},\ \href {\doibase 10.1103/PhysRevLett.84.4457} {\bibfield
  {journal} {\bibinfo  {journal} {Phys. Rev. Lett.}\ }\textbf {\bibinfo
  {volume} {84}},\ \bibinfo {pages} {4457} (\bibinfo {year}
  {2000})}\BibitemShut {NoStop}%
\bibitem [{\citenamefont {Moessner}\ and\ \citenamefont
  {Sondhi}(2001)}]{Moessner2001}%
  \BibitemOpen
  \bibfield  {author} {\bibinfo {author} {\bibfnamefont {R.}~\bibnamefont
  {Moessner}}\ and\ \bibinfo {author} {\bibfnamefont {S.~L.}\ \bibnamefont
  {Sondhi}},\ }\emph {Ising models of quantum frustration},\ \href {\doibase
  10.1103/PhysRevB.63.224401} {\bibfield  {journal} {\bibinfo  {journal} {Phys.
  Rev. B}\ }\textbf {\bibinfo {volume} {63}},\ \bibinfo {pages} {224401}
  (\bibinfo {year} {2001})}\BibitemShut {NoStop}%
\bibitem [{\citenamefont {Isakov}\ and\ \citenamefont
  {Moessner}(2003)}]{Isakov2003}%
  \BibitemOpen
  \bibfield  {author} {\bibinfo {author} {\bibfnamefont {S.~V.}\ \bibnamefont
  {Isakov}}\ and\ \bibinfo {author} {\bibfnamefont {R.}~\bibnamefont
  {Moessner}},\ }\emph {Interplay of quantum and thermal fluctuations in a
  frustrated magnet},\ \href {\doibase 10.1103/PhysRevB.68.104409} {\bibfield
  {journal} {\bibinfo  {journal} {Phys. Rev. B}\ }\textbf {\bibinfo {volume}
  {68}},\ \bibinfo {pages} {104409} (\bibinfo {year} {2003})}\BibitemShut
  {NoStop}%
\bibitem [{\citenamefont {Koziol}\ \emph {et~al.}(2019)\citenamefont {Koziol},
  \citenamefont {Fey}, \citenamefont {Kapfer},\ and\ \citenamefont
  {Schmidt}}]{Koziol2019}%
  \BibitemOpen
  \bibfield  {author} {\bibinfo {author} {\bibfnamefont {J.}~\bibnamefont
  {Koziol}}, \bibinfo {author} {\bibfnamefont {S.}~\bibnamefont {Fey}},
  \bibinfo {author} {\bibfnamefont {S.~C.}\ \bibnamefont {Kapfer}}, \ and\
  \bibinfo {author} {\bibfnamefont {K.~P.}\ \bibnamefont {Schmidt}},\ }\emph
  {{Quantum criticality of the transverse-field Ising model with long-range
  interactions on triangular-lattice cylinders}},\ \href {\doibase
  10.1103/PhysRevB.100.144411} {\bibfield  {journal} {\bibinfo  {journal}
  {Phys. Rev. B}\ }\textbf {\bibinfo {volume} {100}},\ \bibinfo {pages}
  {144411} (\bibinfo {year} {2019})}\BibitemShut {NoStop}%
\bibitem [{\citenamefont {Madjarov}\ \emph {et~al.}(2019)\citenamefont
  {Madjarov}, \citenamefont {Cooper}, \citenamefont {Shaw}, \citenamefont
  {Covey}, \citenamefont {Schkolnik}, \citenamefont {Yoon}, \citenamefont
  {Williams},\ and\ \citenamefont {Endres}}]{Madjarov2019}%
  \BibitemOpen
  \bibfield  {author} {\bibinfo {author} {\bibfnamefont {I.~S.}\ \bibnamefont
  {Madjarov}}, \bibinfo {author} {\bibfnamefont {A.}~\bibnamefont {Cooper}},
  \bibinfo {author} {\bibfnamefont {A.~L.}\ \bibnamefont {Shaw}}, \bibinfo
  {author} {\bibfnamefont {J.~P.}\ \bibnamefont {Covey}}, \bibinfo {author}
  {\bibfnamefont {V.}~\bibnamefont {Schkolnik}}, \bibinfo {author}
  {\bibfnamefont {T.~H.}\ \bibnamefont {Yoon}}, \bibinfo {author}
  {\bibfnamefont {J.~R.}\ \bibnamefont {Williams}}, \ and\ \bibinfo {author}
  {\bibfnamefont {M.}~\bibnamefont {Endres}},\ }\emph {An Atomic-Array Optical
  Clock with Single-Atom Readout},\ \href {\doibase 10.1103/PhysRevX.9.041052}
  {\bibfield  {journal} {\bibinfo  {journal} {Phys. Rev. X}\ }\textbf {\bibinfo
  {volume} {9}},\ \bibinfo {pages} {041052} (\bibinfo {year}
  {2019})}\BibitemShut {NoStop}%
\bibitem [{\citenamefont {Norcia}\ \emph {et~al.}(2019)\citenamefont {Norcia},
  \citenamefont {Young}, \citenamefont {Eckner}, \citenamefont {Oelker},
  \citenamefont {Ye},\ and\ \citenamefont {Kaufman}}]{Norcia2019}%
  \BibitemOpen
  \bibfield  {author} {\bibinfo {author} {\bibfnamefont {M.~A.}\ \bibnamefont
  {Norcia}}, \bibinfo {author} {\bibfnamefont {A.~W.}\ \bibnamefont {Young}},
  \bibinfo {author} {\bibfnamefont {W.~J.}\ \bibnamefont {Eckner}}, \bibinfo
  {author} {\bibfnamefont {E.}~\bibnamefont {Oelker}}, \bibinfo {author}
  {\bibfnamefont {J.}~\bibnamefont {Ye}}, \ and\ \bibinfo {author}
  {\bibfnamefont {A.~M.}\ \bibnamefont {Kaufman}},\ }\emph {Seconds-scale
  coherence on an optical clock transition in a tweezer array},\ \href
  {\doibase 10.1126/science.aay0644} {\bibfield  {journal} {\bibinfo  {journal}
  {Science}\ }\textbf {\bibinfo {volume} {366}},\ \bibinfo {pages} {93}
  (\bibinfo {year} {2019})}\BibitemShut {NoStop}%
\bibitem [{\citenamefont {Saffman}\ \emph {et~al.}(2010)\citenamefont
  {Saffman}, \citenamefont {Walker},\ and\ \citenamefont
  {M\o{}lmer}}]{Saffman2010}%
  \BibitemOpen
  \bibfield  {author} {\bibinfo {author} {\bibfnamefont {M.}~\bibnamefont
  {Saffman}}, \bibinfo {author} {\bibfnamefont {T.~G.}\ \bibnamefont {Walker}},
  \ and\ \bibinfo {author} {\bibfnamefont {K.}~\bibnamefont {M\o{}lmer}},\
  }\emph {Quantum information with Rydberg atoms},\ \href {\doibase
  10.1103/RevModPhys.82.2313} {\bibfield  {journal} {\bibinfo  {journal} {Rev.
  Mod. Phys.}\ }\textbf {\bibinfo {volume} {82}},\ \bibinfo {pages} {2313}
  (\bibinfo {year} {2010})}\BibitemShut {NoStop}%
\bibitem [{\citenamefont {Henriet}\ \emph {et~al.}(2020)\citenamefont
  {Henriet}, \citenamefont {B{\'{e}}guin}, \citenamefont {Signoles},
  \citenamefont {Lahaye}, \citenamefont {Browaeys}, \citenamefont {Reymond},\
  and\ \citenamefont {Jurczak}}]{Henriet2020b}%
  \BibitemOpen
  \bibfield  {author} {\bibinfo {author} {\bibfnamefont {L.}~\bibnamefont
  {Henriet}}, \bibinfo {author} {\bibfnamefont {L.}~\bibnamefont
  {B{\'{e}}guin}}, \bibinfo {author} {\bibfnamefont {A.}~\bibnamefont
  {Signoles}}, \bibinfo {author} {\bibfnamefont {T.}~\bibnamefont {Lahaye}},
  \bibinfo {author} {\bibfnamefont {A.}~\bibnamefont {Browaeys}}, \bibinfo
  {author} {\bibfnamefont {G.-O.}\ \bibnamefont {Reymond}}, \ and\ \bibinfo
  {author} {\bibfnamefont {C.}~\bibnamefont {Jurczak}},\ }\emph {Quantum
  computing with neutral atoms},\ \href {\doibase 10.22331/q-2020-09-21-327}
  {\bibfield  {journal} {\bibinfo  {journal} {{Quantum}}\ }\textbf {\bibinfo
  {volume} {4}},\ \bibinfo {pages} {327} (\bibinfo {year} {2020})}\BibitemShut
  {NoStop}%
\bibitem [{\citenamefont {Morgado}\ and\ \citenamefont
  {Whitlock}(2020)}]{Whitlock2020}%
  \BibitemOpen
  \bibfield  {author} {\bibinfo {author} {\bibfnamefont {M.}~\bibnamefont
  {Morgado}}\ and\ \bibinfo {author} {\bibfnamefont {S.}~\bibnamefont
  {Whitlock}},\ }\emph {Quantum simulation and computing with Rydberg qubits},\
  \href {https://arxiv.org/abs/2011.03031} {\bibfield  {journal} {\bibinfo
  {journal} {arXiv:2011.03031}\ } (\bibinfo {year} {2020})}\BibitemShut
  {NoStop}%
\bibitem [{\citenamefont {Barredo}\ \emph {et~al.}(2018)\citenamefont
  {Barredo}, \citenamefont {Lienhard}, \citenamefont {de~L\'es\'eleuc},
  \citenamefont {Lahaye},\ and\ \citenamefont {Browaeys}}]{Barredo2018}%
  \BibitemOpen
  \bibfield  {author} {\bibinfo {author} {\bibfnamefont {D.}~\bibnamefont
  {Barredo}}, \bibinfo {author} {\bibfnamefont {V.}~\bibnamefont {Lienhard}},
  \bibinfo {author} {\bibfnamefont {S.}~\bibnamefont {de~L\'es\'eleuc}},
  \bibinfo {author} {\bibfnamefont {T.}~\bibnamefont {Lahaye}}, \ and\ \bibinfo
  {author} {\bibfnamefont {A.}~\bibnamefont {Browaeys}},\ }\emph {Synthetic
  three-dimensional atomic structures assembled atom by atom},\ \href {\doibase
  10.1038/s41586-018-0450-2} {\bibfield  {journal} {\bibinfo  {journal}
  {Nature}\ }\textbf {\bibinfo {volume} {561}},\ \bibinfo {pages} {79}
  (\bibinfo {year} {2018})}\BibitemShut {NoStop}%
\bibitem [{\citenamefont {de~L\'es\'eleuc}\ \emph {et~al.}(2018)\citenamefont
  {de~L\'es\'eleuc}, \citenamefont {Barredo}, \citenamefont {Lienhard},
  \citenamefont {Browaeys},\ and\ \citenamefont {Lahaye}}]{Leseleuc2018PRA}%
  \BibitemOpen
  \bibfield  {author} {\bibinfo {author} {\bibfnamefont {S.}~\bibnamefont
  {de~L\'es\'eleuc}}, \bibinfo {author} {\bibfnamefont {D.}~\bibnamefont
  {Barredo}}, \bibinfo {author} {\bibfnamefont {V.}~\bibnamefont {Lienhard}},
  \bibinfo {author} {\bibfnamefont {A.}~\bibnamefont {Browaeys}}, \ and\
  \bibinfo {author} {\bibfnamefont {T.}~\bibnamefont {Lahaye}},\ }\emph
  {{Analysis of imperfections in the coherent optical excitation of single
  atoms to Rydberg states}},\ \href {\doibase 10.1103/PhysRevA.97.053803}
  {\bibfield  {journal} {\bibinfo  {journal} {Phys. Rev. A}\ }\textbf {\bibinfo
  {volume} {97}},\ \bibinfo {pages} {053803} (\bibinfo {year}
  {2018})}\BibitemShut {NoStop}%
\bibitem [{\citenamefont {Schollw\"ock}(2011)}]{Schollwoeck2011}%
  \BibitemOpen
  \bibfield  {author} {\bibinfo {author} {\bibfnamefont {U.}~\bibnamefont
  {Schollw\"ock}},\ }\emph {The density-matrix renormalization group in the age
  of matrix product states},\ \href {\doibase
  https://doi.org/10.1016/j.aop.2010.09.012} {\bibfield  {journal} {\bibinfo
  {journal} {Ann. Phys. (NY)}\ }\textbf {\bibinfo {volume} {326}},\ \bibinfo
  {pages} {96 } (\bibinfo {year} {2011})}\BibitemShut {NoStop}%
\bibitem [{\citenamefont {Paeckel}\ \emph {et~al.}(2019)\citenamefont
  {Paeckel}, \citenamefont {K\"ohler}, \citenamefont {Swoboda}, \citenamefont
  {Manmana}, \citenamefont {Schollw\"ock},\ and\ \citenamefont
  {Hubig}}]{Paeckel2019}%
  \BibitemOpen
  \bibfield  {author} {\bibinfo {author} {\bibfnamefont {S.}~\bibnamefont
  {Paeckel}}, \bibinfo {author} {\bibfnamefont {T.}~\bibnamefont {K\"ohler}},
  \bibinfo {author} {\bibfnamefont {A.}~\bibnamefont {Swoboda}}, \bibinfo
  {author} {\bibfnamefont {S.~R.}\ \bibnamefont {Manmana}}, \bibinfo {author}
  {\bibfnamefont {U.}~\bibnamefont {Schollw\"ock}}, \ and\ \bibinfo {author}
  {\bibfnamefont {C.}~\bibnamefont {Hubig}},\ }\emph {Time-evolution methods
  for matrix-product states},\ \href {\doibase
  https://doi.org/10.1016/j.aop.2019.167998} {\bibfield  {journal} {\bibinfo
  {journal} {Ann. Phys. (NY)}\ }\textbf {\bibinfo {volume} {411}},\ \bibinfo
  {pages} {167998} (\bibinfo {year} {2019})}\BibitemShut {NoStop}%
\bibitem [{\citenamefont {Stoudenmire}\ and\ \citenamefont
  {White}(2012)}]{Stoudenmire2012}%
  \BibitemOpen
  \bibfield  {author} {\bibinfo {author} {\bibfnamefont {E.~M.}\ \bibnamefont
  {Stoudenmire}}\ and\ \bibinfo {author} {\bibfnamefont {S.~R.}\ \bibnamefont
  {White}},\ }\emph {Studying Two-Dimensional Systems with the Density Matrix
  Renormalization Group},\ \href {\doibase
  10.1146/annurev-conmatphys-020911-125018} {\bibfield  {journal} {\bibinfo
  {journal} {Annu. Rev. Condens. Matter Phys}\ }\textbf {\bibinfo {volume}
  {3}},\ \bibinfo {pages} {111} (\bibinfo {year} {2012})}\BibitemShut {NoStop}%
\bibitem [{\citenamefont {Eisert}\ \emph {et~al.}(2010)\citenamefont {Eisert},
  \citenamefont {Cramer},\ and\ \citenamefont {Plenio}}]{Eisert10}%
  \BibitemOpen
  \bibfield  {author} {\bibinfo {author} {\bibfnamefont {J.}~\bibnamefont
  {Eisert}}, \bibinfo {author} {\bibfnamefont {M.}~\bibnamefont {Cramer}}, \
  and\ \bibinfo {author} {\bibfnamefont {M.~B.}\ \bibnamefont {Plenio}},\
  }\emph {Colloquium: Area laws for the entanglement entropy},\ \href {\doibase
  10.1103/RevModPhys.82.277} {\bibfield  {journal} {\bibinfo  {journal} {Rev.
  Mod. Phys.}\ }\textbf {\bibinfo {volume} {82}},\ \bibinfo {pages} {277}
  (\bibinfo {year} {2010})}\BibitemShut {NoStop}%
\bibitem [{\citenamefont {Haegeman}\ \emph {et~al.}(2011)\citenamefont
  {Haegeman}, \citenamefont {Cirac}, \citenamefont {Osborne}, \citenamefont
  {Pi\ifmmode~\check{z}\else \v{z}\fi{}orn}, \citenamefont {Verschelde},\ and\
  \citenamefont {Verstraete}}]{Haegman11}%
  \BibitemOpen
  \bibfield  {author} {\bibinfo {author} {\bibfnamefont {J.}~\bibnamefont
  {Haegeman}}, \bibinfo {author} {\bibfnamefont {J.~I.}\ \bibnamefont {Cirac}},
  \bibinfo {author} {\bibfnamefont {T.~J.}\ \bibnamefont {Osborne}}, \bibinfo
  {author} {\bibfnamefont {I.}~\bibnamefont {Pi\ifmmode~\check{z}\else
  \v{z}\fi{}orn}}, \bibinfo {author} {\bibfnamefont {H.}~\bibnamefont
  {Verschelde}}, \ and\ \bibinfo {author} {\bibfnamefont {F.}~\bibnamefont
  {Verstraete}},\ }\emph {Time-Dependent Variational Principle for Quantum
  Lattices},\ \href {\doibase 10.1103/PhysRevLett.107.070601} {\bibfield
  {journal} {\bibinfo  {journal} {Phys. Rev. Lett.}\ }\textbf {\bibinfo
  {volume} {107}},\ \bibinfo {pages} {070601} (\bibinfo {year}
  {2011})}\BibitemShut {NoStop}%
\bibitem [{\citenamefont {Haegeman}\ \emph {et~al.}(2016)\citenamefont
  {Haegeman}, \citenamefont {Lubich}, \citenamefont {Oseledets}, \citenamefont
  {Vandereycken},\ and\ \citenamefont {Verstraete}}]{Haegman16}%
  \BibitemOpen
  \bibfield  {author} {\bibinfo {author} {\bibfnamefont {J.}~\bibnamefont
  {Haegeman}}, \bibinfo {author} {\bibfnamefont {C.}~\bibnamefont {Lubich}},
  \bibinfo {author} {\bibfnamefont {I.}~\bibnamefont {Oseledets}}, \bibinfo
  {author} {\bibfnamefont {B.}~\bibnamefont {Vandereycken}}, \ and\ \bibinfo
  {author} {\bibfnamefont {F.}~\bibnamefont {Verstraete}},\ }\emph {Unifying
  time evolution and optimization with matrix product states},\ \href {\doibase
  10.1103/PhysRevB.94.165116} {\bibfield  {journal} {\bibinfo  {journal} {Phys.
  Rev. B}\ }\textbf {\bibinfo {volume} {94}},\ \bibinfo {pages} {165116}
  (\bibinfo {year} {2016})}\BibitemShut {NoStop}%
\bibitem [{\citenamefont {Hauschild}\ and\ \citenamefont
  {Pollmann}(2018)}]{Hauschild18}%
  \BibitemOpen
  \bibfield  {author} {\bibinfo {author} {\bibfnamefont {J.}~\bibnamefont
  {Hauschild}}\ and\ \bibinfo {author} {\bibfnamefont {F.}~\bibnamefont
  {Pollmann}},\ }\emph {{Efficient numerical simulations with Tensor Networks:
  Tensor Network Python (TeNPy)}},\ \href {\doibase
  10.21468/SciPostPhysLectNotes.5} {\bibfield  {journal} {\bibinfo  {journal}
  {SciPost Phys. Lect. Notes}\ ,\ \bibinfo {pages} {5}} (\bibinfo {year}
  {2018})}\BibitemShut {NoStop}%
\bibitem [{\citenamefont {Stoudenmire}\ and\ \citenamefont
  {White}(2010)}]{Stoudenmire2010}%
  \BibitemOpen
  \bibfield  {author} {\bibinfo {author} {\bibfnamefont {E.~M.}\ \bibnamefont
  {Stoudenmire}}\ and\ \bibinfo {author} {\bibfnamefont {S.~R.}\ \bibnamefont
  {White}},\ }\emph {Minimally entangled typical thermal state algorithms},\
  \href {\doibase 10.1088/1367-2630/12/5/055026} {\bibfield  {journal}
  {\bibinfo  {journal} {New J. Phys}\ }\textbf {\bibinfo {volume} {12}},\
  \bibinfo {pages} {055026} (\bibinfo {year} {2010})}\BibitemShut {NoStop}%
\bibitem [{\citenamefont {Marcuzzi}\ \emph {et~al.}(2017)\citenamefont
  {Marcuzzi}, \citenamefont {Min\'a\v{r}}, \citenamefont {Barredo},
  \citenamefont {de~L\'es\'eleuc}, \citenamefont {Labuhn}, \citenamefont
  {Lahaye}, \citenamefont {Browaeys}, \citenamefont {Levi},\ and\ \citenamefont
  {Lesanovsky}}]{Marcuzzi2017}%
  \BibitemOpen
  \bibfield  {author} {\bibinfo {author} {\bibfnamefont {M.}~\bibnamefont
  {Marcuzzi}}, \bibinfo {author} {\bibfnamefont {J.}~\bibnamefont
  {Min\'a\v{r}}}, \bibinfo {author} {\bibfnamefont {D.}~\bibnamefont
  {Barredo}}, \bibinfo {author} {\bibfnamefont {S.}~\bibnamefont
  {de~L\'es\'eleuc}}, \bibinfo {author} {\bibfnamefont {H.}~\bibnamefont
  {Labuhn}}, \bibinfo {author} {\bibfnamefont {T.}~\bibnamefont {Lahaye}},
  \bibinfo {author} {\bibfnamefont {A.}~\bibnamefont {Browaeys}}, \bibinfo
  {author} {\bibfnamefont {E.}~\bibnamefont {Levi}}, \ and\ \bibinfo {author}
  {\bibfnamefont {I.}~\bibnamefont {Lesanovsky}},\ }\emph {Facilitation
  Dynamics and Localization Phenomena in Rydberg Lattice Gases with Position
  Disorder},\ \href {\doibase 10.1103/PhysRevLett.118.063606} {\bibfield
  {journal} {\bibinfo  {journal} {Phys. Rev. Lett.}\ }\textbf {\bibinfo
  {volume} {118}},\ \bibinfo {pages} {063606} (\bibinfo {year}
  {2017})}\BibitemShut {NoStop}%
\bibitem [{\citenamefont {Binder}(1981)}]{Binder81}%
  \BibitemOpen
  \bibfield  {author} {\bibinfo {author} {\bibfnamefont {K.}~\bibnamefont
  {Binder}},\ }\emph {{Finite size scaling analysis of ising model block
  distribution functions}},\ \href {\doibase 10.1007/BF01293604} {\bibfield
  {journal} {\bibinfo  {journal} {Z. Phys. B: Condens. Matter}\ }\textbf
  {\bibinfo {volume} {43}},\ \bibinfo {pages} {119} (\bibinfo {year}
  {1981})}\BibitemShut {NoStop}%
\bibitem [{\citenamefont {Onsager}(1944)}]{Onsager1944}%
  \BibitemOpen
  \bibfield  {author} {\bibinfo {author} {\bibfnamefont {L.}~\bibnamefont
  {Onsager}},\ }\emph {{Crystal Statistics. I. A Two-Dimensional Model with an
  Order-Disorder Transition}},\ \href {\doibase 10.1103/PhysRev.65.117}
  {\bibfield  {journal} {\bibinfo  {journal} {Phys. Rev.}\ }\textbf {\bibinfo
  {volume} {65}},\ \bibinfo {pages} {117} (\bibinfo {year} {1944})}\BibitemShut
  {NoStop}%
\end{thebibliography}%

\section*{Supplementary Information} 

\subsection{Experimental set-up} \label{SM:expSetup}

Our experimental setup, described in detail in Ref.~\citenum{Barredo2018}, 
is based on arrays of single atoms trapped in optical 
tweezers. 
In order to create defect-free arrays consisting of $N$ atoms, 
we begin with a stochastically 
loaded array containing $>2N$ traps and then, using a single, 
moving optical tweezer trap, rearrange the atoms into the 
desired configuration. Optimisation of the rearrangement 
algorithms and the introduction of multiple cycles \cite{Schymik2020} have 
enabled us to create arbitrary arrays with up to 200 atoms. 

Following the rearrangement, the atoms are optically pumped 
into the ground state $\ket{\downarrow} = \ket{5S_{1/2}, F = 2, m_F = 2}$ 
in the presence of a static magnetic field of 7~G with $99.9 \%$ efficiency. 
The tweezers are then switched off for the duration of the sweeps. 
We use the Rydberg state $\ket{\uparrow}=\ket{75S_{1/2}, m_J = 1/2}$ 
with a lifetime $\tau=175\;\mu{\rm s}$, and van der Waals coefficient $C_6/h=1947\;{\rm GHz}\cdot\mu{\rm m}^6$.
To excite the atoms from $\ket{\downarrow}$ to $\ket{\uparrow}$, 
we drive a two-photon transition using counter-propagating 
laser beams with wavelengths 420~nm and 1013~nm, via the 
intermediate state $\ket{6P_{3/2}, F=3, m_{\rm{F}}=3}$ with a 
lifetime of 113~ns. We use two Ti:Sapphire lasers (M-Squared), 
because of their intrinsic low phase noise at high frequency\cite{Leseleuc2018PRA}. 
One, operating at 840~nm, is frequency doubled to 420~nm with 
up to 2~W, and the second, at 1013~nm, seeds an AzurLight Systems 
fibre amplifier delivering up to 10~W. The 420~nm light is fibre 
coupled to the experiment, whilst the 1013~nm light is free-space. 
The $1/e^2$ radii and maximum powers of the lasers at the position of the 
atoms are $w_{420}=250~\mu$m, $P_{420}=350$~mW, and 
$w_{1013}=130~\mu$m, $P_{1013}=5$~W, allowing for Rabi 
frequencies of up to $\Omega_{420}=2\pi \times 200$~MHz 
and $\Omega_{1013}=2\pi\times50$~MHz. To limit the spontaneous 
emission, the lasers are detuned by $ 2 \pi \times 700 $~MHz 
from the intermediate state, resulting in a maximum effective 
Rabi frequency of $\Omega=2\pi \times 7$~MHz.

To detect the state of the atoms the tweezers are turned back on, recapturing atoms 
in $\ket{\downarrow}$, which we then image. The atoms in $\ket{\uparrow}$ are 
repelled from the tweezers and hence are not imaged. This detection method features 
imperfections, described in detail in Ref.~\citenum{Leseleuc2018PRA}, 
which result in two types of errors. 
Firstly a false positive, with probability $\varepsilon$, 
where a ground-state atom is lost and misidentified as 
a Rydberg atom, and secondly a false negative, with probability 
$\varepsilon^{\prime}$, where a Rydberg atom decays quickly, 
is recaptured and hence, misidentified as a ground-state atom.
The results presented in the main text have $\varepsilon= 1\%$ and $\varepsilon^{\prime} = 3\%$.

\subsection{Matrix Product States}\label{SM:MPS}
In order to verify the experimental results we compare to numerical simulations. 
For the large atom numbers used in this work Exact Diagonalisation 
methods are too expensive and so we employ matrix product state (MPS) 
methods as described in this section.
\subsubsection{Method}
\label{SM:mpsMethod}

Matrix product state methods are well documented in the 
literature~\cite{Schollwoeck2011,Paeckel2019}, so we focus here 
on the aspects important for the discussion
in this paper. 

MPS methods approximate the physical state by a linear network of $N$ 
tensors, one at each site. Initially developed for one-dimensional many body 
systems, the method is also successful on two-dimensional systems of limited width~\cite{Stoudenmire2012}. 
In our case the ${L \times L}$ square lattice is mapped 
row-by-row to a linear tensor network, for the triangular arrays the mapping 
is similar except that the rows have variable width.
The MPS approximation is controlled by the bond dimension $\chi$, which 
limits the maximum amount of entanglement between consecutive subsystems.
An MPS with bond dimension $\chi$ can exhibit at most an entanglement entropy of $S=\ln(\chi)$ 
for a bipartition into two connected chains.

For ground state wavefunctions of gapped, local Hamiltonians one expects the entanglement entropy of subsystems to follow
an 
area law~\cite{Eisert10}. This means that the entanglement entropy between two subsystems is proportional to the 
length (or area) of the perimeter of the subsystems, and not proportional to their volume. When considering a time-dependent 
situation as we do here, the entanglement structure can become more complicated, and, for example, change to a volume law if
long times after a quench are studied. Since we typically fix the duration of our sweeps while scaling up the system size, 
we are effectively still dealing with an area-law situation. Applying these considerations to the square and the triangular lattice 
arrays, we are simulating systems with a maximal entanglement entropy between subsystems proportional to the {\em width} of 
the clusters, i.e.~$S \sim L$ for the square arrays. This translates into a scaling of the required bond dimension $\chi \sim \exp[\alpha L]$
for simulations of constant accuracy. Based on this scaling we are able to reach widths of $L=10$ for the square lattice and a similar
width for the triangular array ($N=108$) with a bond dimension of up to $\chi=512$, see discussion below.

For the unitary time evolution we use the time-dependent variation 
principle (TDVP) \cite{Haegman11, Haegman16} which works by projecting the Hamiltonian onto 
the state's tangent space and then applying the time evolution operator for a small time-step.
At each step we can either use a single-site or a two-site algorithm, 
the latter of which allows for growth of the MPS' bond dimension.
The simulation starts in the initial product state which evolves in time according to the Hamiltonian \eqref{eq:tfiRyd}, 
at first using the two-site variant until the bond dimension is saturated to a 
certain chosen value $\chi$, and then we switch to the single-site variant for a significant performance increase.

For every system size these simulations are repeated for 30 to 70 
different realisations of random lattice positions,
and include other experimental uncertainties, described in Sec.~\ref{SM:imp}.
We use the implementation of TDVP provided by the python package \textit{TeNPy} \cite{Hauschild18}.
The computationally most expensive simulations for the $10\times10$ square with a bond dimension of $\chi = 512$ take up to 14 days for a single disorder instance. We use computer clusters to simultaneously simulate around 50 disorder instances.

\subsubsection{Sampling}
\label{SM:mpsSampling}
The experiment delivers snapshots of the state of the system 
projected to the Rydberg Fock space. From this data, one can determine 
local densities, density correlation functions, and sublattice histograms. 
These quantities can also be obtained from Exact Diagonalization, as well 
as the classical and quantum Monte Carlo algorithms. 
With the MPS method local densities and 
correlations can also be obtained
easily by tensor contractions. However, the sampling of the Fock space snapshots is not so simple. 
While the weight of a given configuration can be efficiently 
obtained by contracting the MPS network with the snapshot configuration
as the input on the physical legs, it is less straightforward to find the 
most probable configurations out of the $2^{N}$ possible states.

We devised an algorithm to generate statistically independent snapshots of a given MPS wave function:
\begin{enumerate}

	\item Pick an arbitrary site $i$ among the $n$ unprojected sites of the normalised MPS, and determine the 
	diagonal elements of the single site reduced density matrix $p_g^{(i)}$ 
	and $p_e^{(i)}$ (for the ground state ($\ket{\downarrow}$) and excited state ($\ket{\uparrow}$) respectively) which sum up to one.
	\item Draw a random number $r\in [0,1]$ and select $g$ if $r<p_g^{(i)}$ or else select $e$.
	Apply the single site projector for the selected subspace, effectively fixing the physical leg of that site's tensor.
	\item The remaining MPS with $n-1$ unprojected sites is normalised to the value of $p_{g|e}^{(i)}$, depending on the randomly selected subspace in the previous step. Thus we divide the MPS by that value to normalise it to 1.
	\item Repeat the process until there are no unprojected sites remaining.
\end{enumerate}

It does not matter whether the sites to be projected are picked randomly or in sequence, however the latter provides 
the opportunity to optimise the generation of thousands of snapshots
by pre-contracting the tensor network. We usually generate 1000 to 10000 snapshots for each data point shown. 
The algorithm presented here is related to the ``collapse'' step in ``Minimally entangled typical thermal state'' MPS algorithms
for finite temperature simulations\cite{Stoudenmire2010}.

\begin{figure}
\includegraphics[width=8.5 cm]{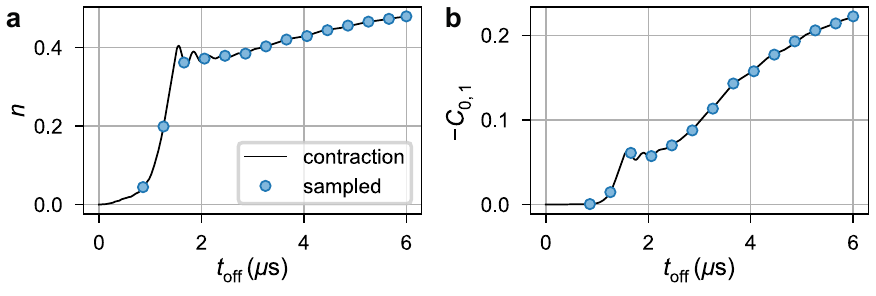}
\caption{\textbf{MPS Sampling.} 
	 \textbf{a}, Average Rydberg density and \textbf{b} nearest-neighbour correlation function during the MPS state dynamics on the ${10\times10}$ square lattice. 
	 The black lines show the observables computed from standard tensor contraction, the blue dots show the corresponding sample average of 1000 generated snapshots. 
	 }
\label{fig:SMmpssampling}
\end{figure}

In \figref{fig:SMmpssampling} we compare the average Rydberg density $n=\sum_i \langle n_i\rangle/N$ and the correlator $C_{0, 1}$ during the time-evolution on a ${10\times10}$ square lattice computed by standard tensor contractions (black curves) to their value from a sample average of 1000 snapshots, generated with the algorithm discussed above (blue dots).
The perfect agreement demonstrates that the devised algorithm successfully samples the MPS wave function.

\subsubsection{Bond dimension $\chi$ dependence}
\label{SM:mpsScaling}
\begin{figure}
\includegraphics[width=8.5 cm]{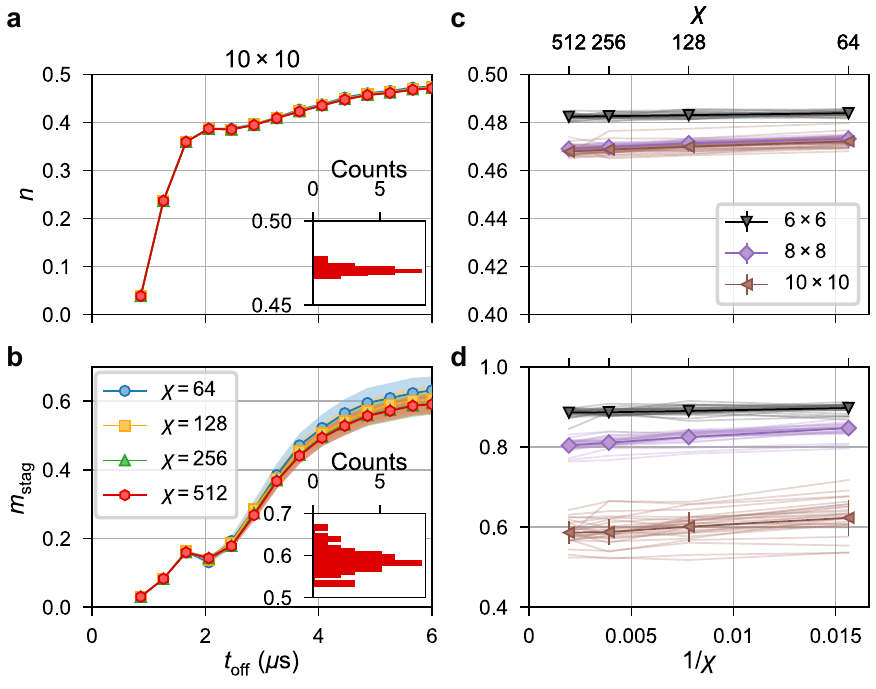}
\caption{\textbf{Scaling of observables with MPS bond dimension $\chi$.} 
	\textbf{a,} Rydberg density and \textbf{b,} order parameter during the MPS state dynamics, including experimental imperfections, for different $\chi$ on the $10\times10$ square lattice. The insets show the distribution due to the multiple $U_{ij}$ disorder realisations, at final $t_{\rm off} = 6 \ \mu {\rm s}$ for $\chi=256$.  
	Scaling of \textbf{c}, $n$ and \textbf{d}, $m_{\rm{stag}}$  with $\chi$ at the end of the state preparation protocol for different system sizes. The lightly coloured lines show the multiple disorder instances.} 
\label{fig:SMchiscaling}
\end{figure}

Due to the aforementioned scaling of the computational complexity of the problem with $\chi$, a trade-off between accuracy and available computational resources has to be achieved. In \figref{fig:SMchiscaling} we analyse the scaling of relevant observables used in this paper with the bond dimension $\chi$, including experimental imperfections. Both the average Rydberg density $n$ (\figref{fig:SMchiscaling}a) and the order parameter $m_{\rm stag}$ (\figref{fig:SMchiscaling}b) show only a very weak dependence on $\chi$ for small times $t_{\rm off}$. 
As expected, for larger times, the variation with $\chi$ becomes stronger, in particular for $m_{\rm stag}$.
Nevertheless, this variation remains small even at the end of the sweep  and, notably, the uncertainty from different interaction disorder instances, shown by the shaded regions, is comparable to the variation between the lowest and largest bond dimensions. 

In \figref{fig:SMchiscaling}c,d, we plot $n$ and $m_{\rm stag}$ at the end of the sweep versus $1/\chi$ for different system sizes. For $L=6$, simulations with up to $\chi=512$ reveal a nice convergence for these observables, and the larger systems show only a slight dependence on $\chi$. The variation resulting from $\chi$ is again smaller than that of the random $U_{ij}$, illustrated by the lighter lines. 

To summarise, our MPS simulations for systems up to $L=10$ are reliable to characterise and benchmark the experimental results, although they might not be fully converged in $\chi$ with regard
to other observables, such as the entanglement entropy. 
Finally, we want to mention that reliable simulations of the dynamics for the largest experimental results achieved in this paper ($L=14$) seem to be out of reach with currently available computational hardware, because of the exponential scaling of $\chi$ with linear system size $L$.  

\subsection{Benchmarking the platform}\label{SM:benchmark}

In this section we discuss experimental imperfections, 
and systematically compare the results of the experiment with numerical simulations. 
We first investigate the coherence of the laser excitation for a single atom, and then introduce 
the imperfections relevant to the many-atom case.

\subsubsection{Coherence of single-atom laser excitation}
\label{SM:rabi}

\begin{figure}
	\includegraphics[width=8.5 cm]{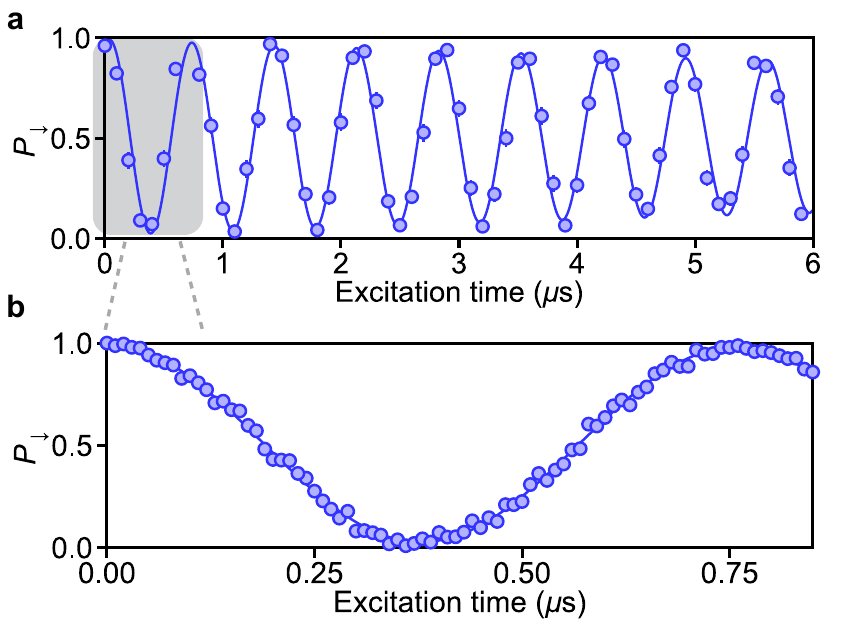}
	\caption{{\bf Testing the coherence of the Rydberg-excitation on a single atom.} 
		\textbf{a}, Rabi oscillations showing the probability of measuring the atom 
		in  $\ket{\downarrow}$ as a function of the excitation time. 
		The line is a fit to the data by the function  $Ae^{-\Gamma t}\cos(\Omega t)+B$, 
		yielding $\Gamma=0.04(1) \mu \text{s}^{-1}$, $\Omega=2\pi \times 1.32(1)$~MHz, 
		$A=0.975(6)$, and $B=0.507(3)$. 
		\textbf{b}, High resolution measurement of the first period of the oscillation. 
		Error bars are statistical and often smaller than marker size.}
	\label{fig:SMRabi}
\end{figure}

To investigate the efficiency and the coherence of the Rydberg excitation on a single atom, 
we drive Rabi oscillations between the $\ket{\downarrow}$ and $\ket{\uparrow}$ states. The results are shown in Fig.~\ref{fig:SMRabi}a, 
where the oscillations are fitted by a damped sine $Ae^{-\Gamma t}\cos(\Omega t)+B$. 
We obtain a decay rate $\Gamma=0.04(1) \mu \text{s}^{-1}$ due to spontaneous 
emission from the intermediate state and to the shot-to-shot fluctuations 
of the laser power ($\sim1\%$). In Fig.~\ref{fig:SMRabi}b, 
we show a magnification of the first period, from which we measure a 
97.3\% probability of transferring a single atom to the $\ket{\uparrow}$ 
state with a $\pi$-pulse. For this experiment, following the methods outlined 
in Ref.~\citenum{Leseleuc2018PRA}, we measure $\varepsilon = 0.4 \%$ and 
infer from simulation using the experimental parameters $\varepsilon^{'} = 1.9 \%$. 
When accounting for these detection errors, we extract a probability of 99.1\% 
to transfer a single atom to the $\ket{\uparrow}$ state.

\subsubsection{Benchmarking the $4 \times 4$ array}\label{SM:benchmarking4x4}

\begin{figure}
\includegraphics[width = 8.5 cm]{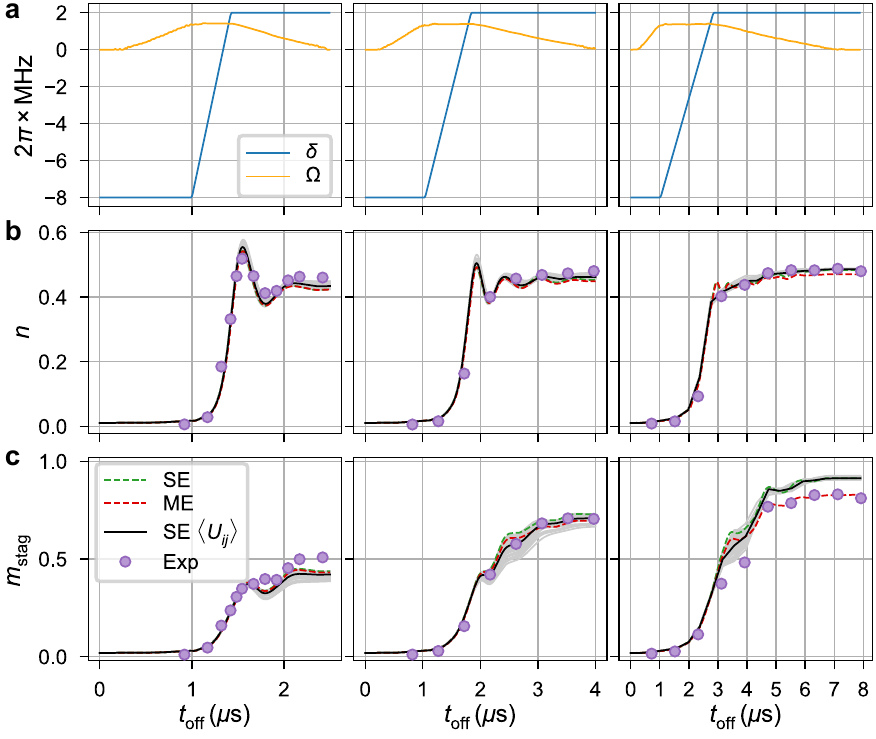}
\caption{\textbf{Benchmarking multiple sweeps on the $4\times4$ array} Time evolution of \textbf{a}, sweep shape, \textbf{b}, Rydberg density, and \textbf{c}, staggered magnetisation for three distinct sweeps (of durations $2.5~\mu$s, $4~\mu$s and $8~\mu$s) on the $4\times4$ array.
In (\textbf{b}, \textbf{c}) experimental data is shown in purple circles, the green (red) dashed line shows solutions to the Schr\"odinger (Lindblad master) equation. Solid grey lines show solutions of the Schr\"odinger equation for several random instances of the interaction disorder (see text), the black line is the average over these instances. 
}
\label{fig:SMbenchmark4x4}
\end{figure}

Here we investigate the dynamics of the system for different sweeps on the $4\times4$ array. 
This system size can be fully simulated by solving the Schr\"odinger equation (SE), for which we
include the detection errors (here $\varepsilon=1\%$ and $\varepsilon^{\prime}=3\%$). 
We assess the adiabaticity of the drive by performing three sweeps of durations
$2.5~\mu\text{s}$,  $4~\mu\text{s}$ and $8~\mu\text{s}$. 
The parameters of the sweeps, shown in \figref{fig:SMbenchmark4x4}a, are as follows: 
at a detuning $\delta/2\pi = -8$~MHz, $\Omega$ is linearly 
increased from 0 to $\sim 2\pi \times 1.4$~MHz, we then linearly sweep the 
detuning to $\sim 2$~MHz while keeping $\Omega$ constant, 
and finally decrease $\Omega$ to 0. As we drive a two-photon transition, 
the atoms experience a changing light shift as $\Omega$ is swept. 
We counteract this effect by changing $\delta$ accordingly. We use as observables the average Rydberg density $n$ and staggered magnetisation, $m_{\rm{stag}}$.
For all sweeps we observe a good 
agreement between experiment and simulation, especially in the evolution of $n$. As expected, short sweeps 
lead to oscillations in the evolution of $n$ due to a failure to 
adiabatically drive the system. This is confirmed by the value of $m_{\rm{stag}}$ 
at the end of the sweep, $\sim0.5$, which would be 1 if the sweep were adiabatic. 
As the sweep time increases, this oscillatory behaviour 
is reduced and the value of $m_{\rm{stag}}$ increases.

To understand the discrepancies between the SE results and 
the data we consider the potential experimental imperfections relevant for many-atom systems. 
The first is the disorder in atomic positions, which comes from two effects, 
(i) the static disorder in the trap positions (standard deviation $\sim100$~nm), and 
(ii) the finite temperature (10~$\mu$K) of the atoms leading to shot-to-shot 
fluctuations in atom position with a standard deviation of $\sigma_r=170$~nm in 
the plane of the array, and $\sigma_z=1~\mu$m in the transverse direction. 
This results in a spatially correlated, non-gaussian distribution of the interaction energy $U_{ij}$ (see also \figref{fig:SMParameters}e,f)\cite{Marcuzzi2017}. 
This effect is included in simulations by repetitions with randomly assigned atom positions chosen from a Gaussian distribution with the mentioned standard deviations. We run simulations 
of the dynamics and average over many runs of the distribution in 
interaction energies. The runs are shown in Fig.\ref{fig:SMbenchmark4x4} 
in light grey, with the average  SE$\langle U_{ij}\rangle$in black. This disorder has very little influence on the results 
and cannot explain the remaining disagreement. 
Another potential source of imperfection comes from the finite size of the 
1013 nm beam, leading to inhomogeneous $\Omega$ and $\delta$ 
(due to the light-shift of the two-photon transition) across the array, 
however on this small system the inhomogeneity is found to have 
negligible effect on the results. 
The observation that the discrepancy is largest for the longest sweep duration indicates 
that decoherence plays an important role.

To study this effect we solve the Master equation (ME) for which 
the time evolution of the density matrix $\rho(t)$ is described by 
\begin{equation}\label{ME}
\frac{d}{dt}\rho=-\frac{i}{\hbar}[H,\rho]+\mathcal{L}[\rho],
\end{equation}
with a Liouvillian
\begin{equation}
\mathcal{L}[\rho]=\sum_i\frac{\gamma}{2}(2n_i\rho n_i - n_i\rho-\rho n_i)
\end{equation}
and a decoherence rate $\gamma=0.05~\mu$s$^{-1}$ which describes well the single-atom Rabi oscillation, when accounting for the shot-to-shot fluctuations of the laser power.

These results agree closely with the experimental data indicating 
that decoherence at the single particle level is sufficient to describe the system.
We conclude from these studies that there is an optimised sweep duration 
which has to be (i) long enough to cross the gaps as adiabatically as possible and 
(ii) short enough to avoid strong decoherence effects. 
Throughout this paper, we limit the duration of the sweeps to be 
about $6 - 7.5~\mu\text{s}$ to fulfil these two conditions. 

\subsubsection{Effect of imperfections  on larger arrays}
\label{SM:imp}
For larger arrays we use the matrix product state (MPS) method, 
described in detail in Sec.~\ref{SM:MPS} to calculate the dynamics 
of the systems during the sweeps. 

\begin{figure}
\includegraphics[width=8.5 cm]{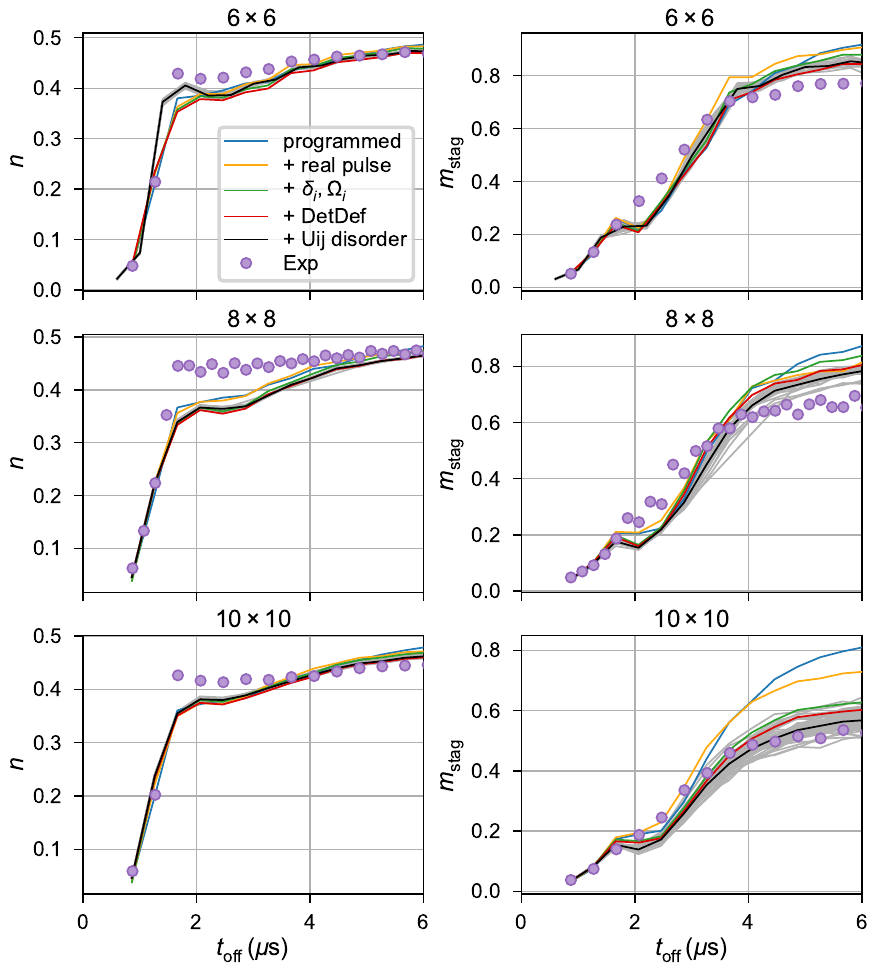}
\caption{\textbf{Effect of experimental imperfections on different system sizes.} The left (right) column shows the average density (order parameter $m_{\rm stag}$). Different lines show successive additions of imperfections on the MPS simulations, starting from the programmed case. Experimental data is shown by circles.}
\label{fig:SMMPSImperfections}
\end{figure}

We first perform the programmed simulation, without experimental imperfections, for 
the $6\times6$, $8\times8$, and $10\times10$ arrays, using the same sweep for each array. Figure \ref{fig:SMMPSImperfections} shows the evolution of $n$ and $m_{\rm{stag}}$, with the experimental data as purple points, and the blue line showing the MPS results. 
We find good agreement over the first $1.5$~$\mu$s after which we observe an overshoot in the experimental results in both $n$ and $m_{\rm stag}$ for all system sizes. We also observe that $m_{\rm stag}$ eventually saturates on the experiment at lower values than the MPS, and that this effect worsens for larger systems.

\begin{figure}
\includegraphics[width=8.5 cm]{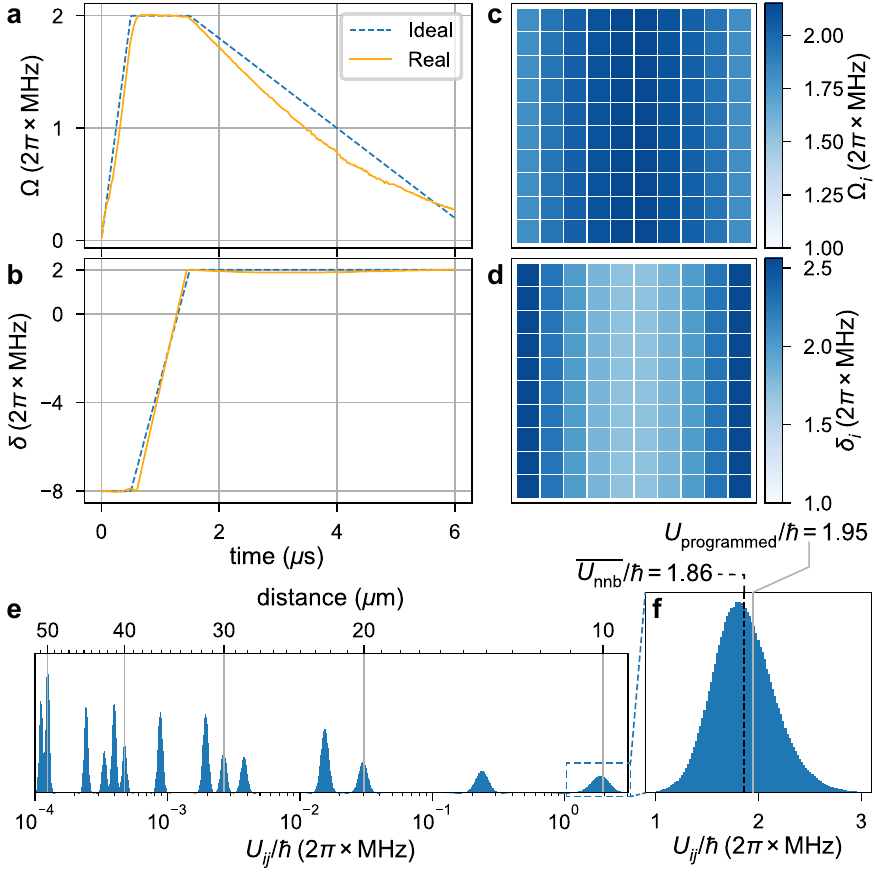}
\caption{\textbf{Experimental parameters for the experiment on the $10 \times 10$ square lattice.} 
\textbf{a-b}, Sweep shape for \textbf{a}, the average Rabi frequency $\Omega$ and \textbf{b}, the average detuning $\delta$ versus time. 
The dashed line shows the proposed protocol, 
the solid line the experimentally obtained parameters.
\textbf{c-d}, Spatial dependence of $\Omega$ (\textbf{c}) and $\delta$ (\textbf{d}) at the maximal values during the protocol.
\textbf{e-f}, Distribution of the Rydberg interactions $U_{ij}$ caused by the 
fluctuations in the atom positions.
\textbf{e}, shows the long-range interactions up to a distance of $\sim 50~\mu {\rm m}$, and \textbf{f}, shows the distribution of the nearest neighbour interactions. The dashed vertical line shows the average nearest-neighbour interaction $U_{\rm nnb}$. The vertical grey line shows, as a reference, the programmed value for non-fluctuating atoms $U_{\rm programmed}$.}
\label{fig:SMParameters}
\end{figure}

We therefore include in the MPS the imperfections discussed in 
the previous section, and illustrated in Fig.~\ref{fig:SMParameters} for the $10\times10$ array, starting from the experimentally measured sweep shape, adding the inhomogeneous fields $\delta_i$, $\Omega_i$, the detection deficiency and, finally, the shot-to-shot fluctuations of the interaction energies caused by the fluctuating atom positions. The results are shown in \figref{fig:SMMPSImperfections}. The imperfections have very little effect on $n$, and are not able to explain the overshoot, in $n$ and $m_{\rm stag}$, at intermediate times. However, at later times we observe better agreement for all system sizes. For the $6\times6$ and the $8\times8$ arrays, each imperfection has a small contribution which tends to decrease the final value of $m_{\rm stag}$. On the $10\times10$ array the largest contribution to the disagreement with the MPS real sweep result is the inhomogeneity of the fields. This is unsurprising as the waist of the 1013~nm laser is comparable to the size of the array.  

\begin{figure}
\includegraphics[width=8.5 cm]{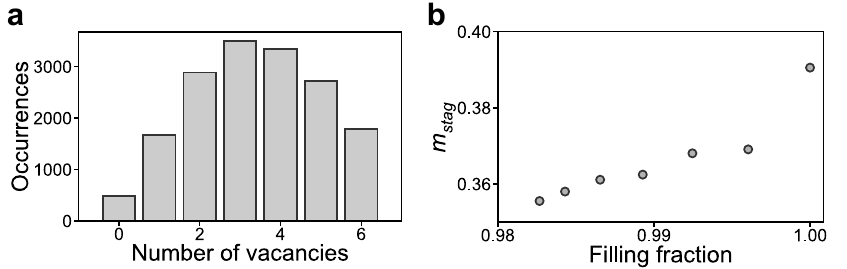}
\caption{\textbf{Effect of vacancies on AF ordering a}, Histogram of the number of defects for the $14 \times 14$ array. 
Out of the $\sim 17000$ experimental realisations shown here, we only 
kept the $\sim 500$ defect-free shots for the results presented in the main text. 
\textbf{b}, $m_{\rm{stag}}$ for different filling fractions. We observe a substantial 
increase in $m_{\rm{stag}}$ for defect-free array experiments, compared to 
$\sim 99 \%$ filled array experiments. }
\label{fig:defects}
\end{figure}	

A final imperfection we consider is the loss of atoms during the rearrangement process, leading to a probability that the prepared array is not perfect. 
After the rearrangement, we take an image of the atoms to 
check the number of vacancies (missing atoms) in the target structure. In Fig.~\ref{fig:defects}a, 
we show the histogram of the number of vacancies for the $14 \times 14$ array, for which $\sim 17000$ experiments have been conducted, with an average filling fraction of $\sim98 \% $. 
Thanks to our high experimental repetition rate of 1--2~Hz, we can post-select the experiments for which the rearrangement 
process was successful. We do this for all results presented in the main text. 
To check the effect of vacancies on the preparation of the antiferromagnetic order, 
we measure $m_{\rm{stag}}$ as a function of the filling fraction. 
To decrease the filling fraction, we increase the number of allowed vacancies. 
The results are shown in~\figref{fig:defects}b. We observe a substantial difference in 
$m_{\rm{stag}}$, between $\sim 99 \% $ filled arrays and defect-free arrays, of approximately $10\%$. This highlights the importance of only considering perfect arrays. 

\subsubsection{Long-term stability of the machine}
\label{SM:stability}

\begin{figure}
	\includegraphics[width=8.5 cm]{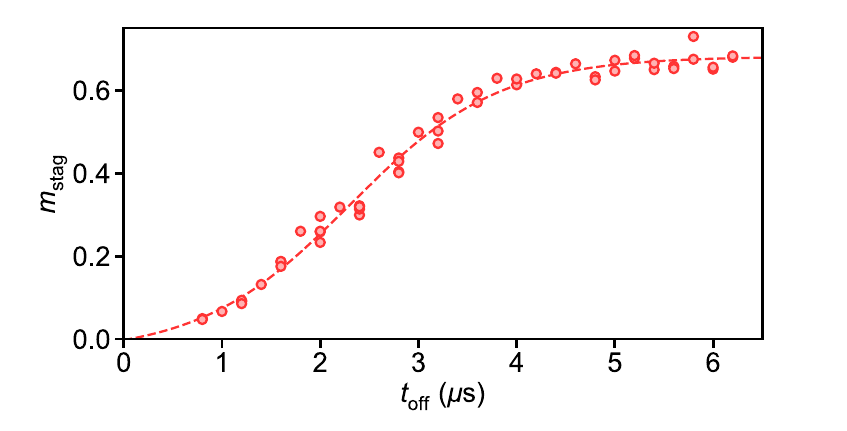}
	\caption{\textbf{Long-term stability of the growth of antiferromagnetic ordering on the $8\times 8$ array.} We show  several measurements for the same parameters, realised typically over 10 hours. We observe a dispersion of the measurements due to long-term drift of the machine. The dashed line is a phenomenological fit to the data. The standard error on the mean is smaller than symbol size.}
	\label{fig:SMstability}
\end{figure}

We now discuss the long-term stability of the machine. We do this analysis on the $8 \times 8$ array by measuring the growth of antiferromagnetic ordering $m_{\rm{stag}}$ throughout the sweep. The results are presented in Fig.~\ref{fig:SMstability}. Each point represents a measurement, with $\sim 300$ snapshots recorded over approximately 20 minutes, giving a standard error on the mean which is smaller than the marker size. We repeat the full curve several times over a duration of about 15 hours and observe a spread of $m_{\rm{stag}}$ throughout the sweep. This is due to long-term instabilities of the experimental machine. These are mainly drifts in beam pointing and power of the Rydberg excitation lasers, which slightly change the path of the sweep in the phase diagram, thus affecting the quality of the prepared antiferromagnetic order. The dashed line is a phenomenological fit to the data and is indicative of the mean evolution of $m_{\rm{stag}}$ regarding the long-term drifts of the machine. We use the standard deviation on the fit for the error bars shown on the final points in \figref{fig2}c,d and \figref{fig4}b.

\subsubsection{Growth of antiferromagnetic order}
\label{SM:AFM}

\begin{figure*}
\includegraphics[width=\textwidth]{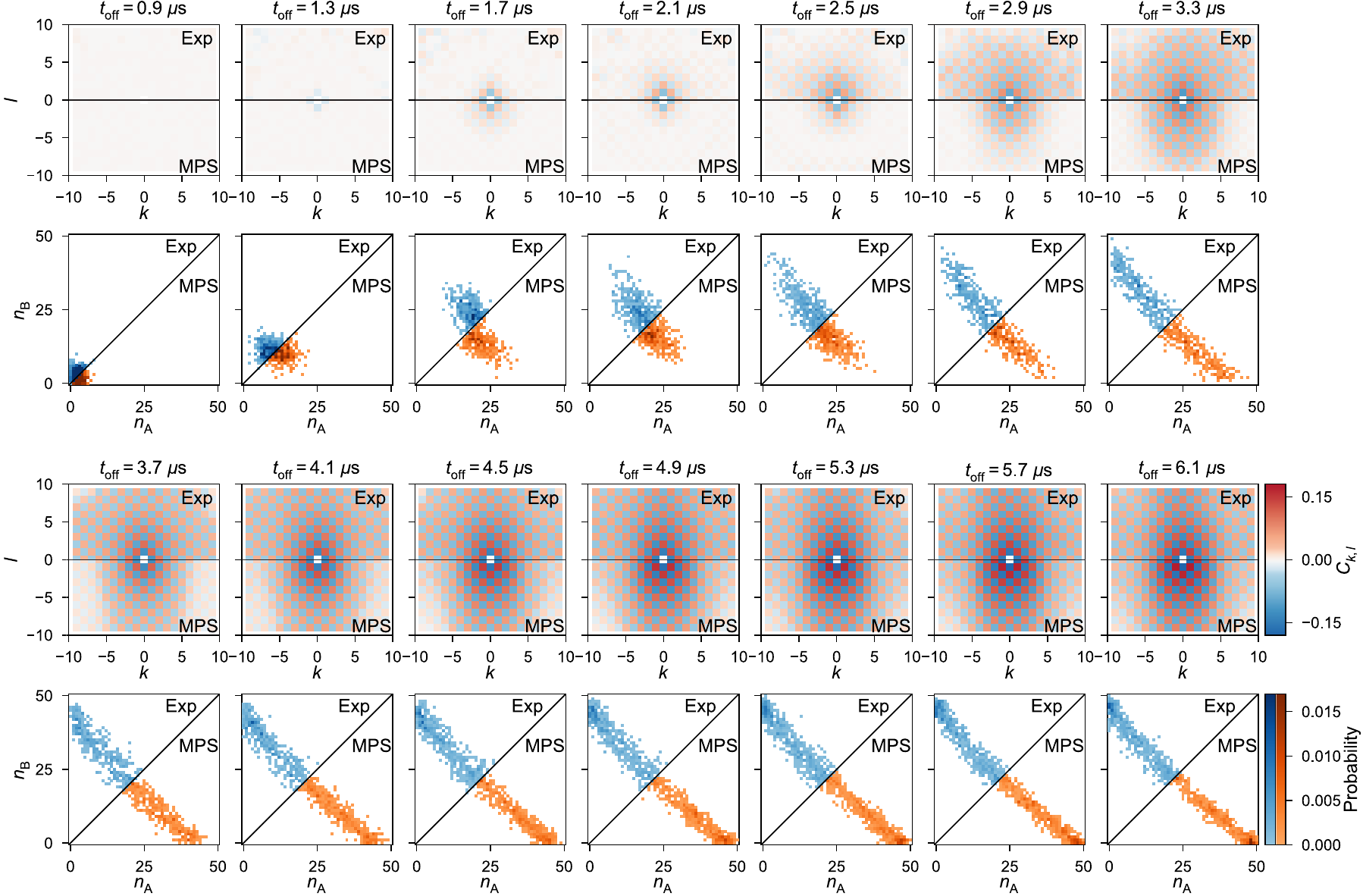}
\caption{\textbf{Growth of antiferromagnetic ordering on a $10\times10$ 
array during the sweep.} Maps of the connected correlations $C_{kl}$ 
and histograms of the staggered magnetisation for different times 
$t_{\rm off}$, defined in Fig.~\ref{fig2}. The upper (lower) halfs of the plots show experimental (MPS) results.}
\label{fig:SMcorr}
\end{figure*}

To further compare the MPS calculations to the experimental results we use the snapshot technique, described in Sec.~\ref{SM:mpsSampling}, to plot the connected correlation maps and the 
$m_{\rm{stag}}$ histograms for each time step in the sweep on the $10\times10$ array. These are shown in \figref{fig:SMcorr}. The $C_{k,l}$ maps show that the correlations are beginning to develop between 1.3 and 1.7~$\mu$s, 
and are growing from nearest-neighbour across the array. By 3.8~$\mu$s 
the expected correlation pattern is filling the entire array, and from then 
on the strength of the correlations continues to increase. 
The corresponding $m_{\rm{stag}}$ histograms show the growth 
in the number of Rydberg excitations as the distribution of points moves 
from $\sim(0,0)$ to higher values of $n_{\rm{A}}+n_{\rm{B}}$. 
As the correlations begin to grow across the array we can see the 
distribution of points stretching along the diagonal and after $3.8~\mu$s 
points start to conglomerate around the corners $(N/2,0)$ and $(0, N/2)$.
The MPS results show qualitative agreement throughout the sweep for both observables, with slight differences appearing at times which correspond to disagreements in \figref{fig:SMMPSImperfections}. This verifies that the dynamical evolution of the atomic system is well approximated by the MPS calculations, including known uncertainties, for all observables. This confirms that we understand and have good control over our platform.

\subsection{Quantum real-time evolution versus classical equilibrium} \label{SM:MCvsMPS}

In this section we give details on the comparison between the quantum real-time evolution and a classical equilibrium description. We also show that the real-time evolution from MPS simulations does not thermalise and cannot be described by a classical equilibrium distribution, similar to the results for the experiment presented in the main text.

\subsubsection{Extracting a classical temperature}
\label{SM:T_hyp}

For a comparison with the classical Hamiltonian (${\Omega = 0}$)
\begin{align}
H_\text{class}[\delta] = \sum_{i < j} U_{ij} n_i n_j - \sum_i \hbar\delta_i n_i\,,
\label{Hclass}
\end{align}
with ${n_i\in\{0,1\}}$, we perform Metropolis Monte Carlo simulations with single spin flip updates. 
Results are averaged over 500 individually equilibrated samples of random atom positions. 
We take interactions up to a Euclidean distance of $5.2 a$, 
or couplings $U_{ij} \gtrapprox 9.9\times 10^{-5}$ MHz into account. 
Note that interactions up to a Euclidean distance of $3.3 a$,
or $U_{ij} \gtrapprox 1.5\times 10^{-3}$ MHz produce nearly 
identical behaviour on the scales considered here. The 
$\delta_i$ are chosen to reflect the experimental setup, shown in \figref{fig:SMParameters}d.

\begin{figure}
	\includegraphics{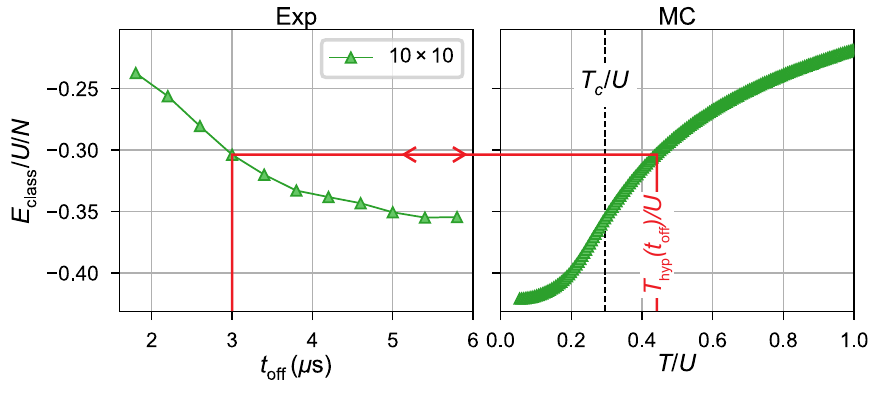}
	\caption{\textbf{Assigning a classical temperature.}
		Classical energy density for the instantaneous state 
		$\left|\Psi(t_{\rm off})\right\rangle$ of the experiment during the last part 
		of the state preparation protocol (left panel), and for the corresponding 
		classical equilibrium system versus temperature $T$ (right panel). 
		Here, $\delta=\delta_f$ for all datasets. 
		We assign a hypothetical temperature $T_{\rm hyp}$ at each time $t_{\rm off}$ 
		by matching the classical energy, as illustrated by the red line.
		In the right panel, $T_c$ denotes the critical temperature 
		in the thermodynamic limit $N\rightarrow\infty$.}
	\label{fig:SMTeff}
\end{figure}

To assign a hypothetical 
temperature $T_{\rm hyp}$ to the dynamical state of the quantum system, we compute the classical energy 
for the instantaneous state $\left| \Psi (t_\text{off}) \right\rangle$,
\begin{equation}
E_{\rm class}(t_{\rm off}) = \left \langle \Psi(t_{\rm off}) | 
H_{\rm class}\left[\delta(t_{\rm off}) \right] | \Psi(t_{\rm off}) \right \rangle
\end{equation}
from the corresponding classical Hamiltonian in Eq.~(\ref{Hclass}).
Since we do not measure the atom positions for each individual experimental snapshot, we assign the interactions $U_{ij} = U_{\rm nnb} / (r_{ij}/a)^6$ to compute $E_{\rm class}$, where $U_{\rm nnb}$ is the atom-position disorder averaged nearest-neighbour interaction (see \figref{fig:SMParameters}f). 
Using classical Monte Carlo we then estimate the 
classical thermodynamic partition function and compute the energy 
expectation value for the same classical Hamiltonian for a temperature $T$ 
\begin{equation}
E_{\rm class}^\mathrm{MC}(T) = \sum_c H_{\rm class}(c) \, \E^{-\beta H_{\rm class}(c)} / Z \ ,
\end{equation}
where we sum over all classical spin configurations $c$ and have defined $\beta = 1/ \left(k_{\rm{B}} T \right)$ and 
$Z = \sum_c \E^{-\beta H_{\rm class}(c)}$. 
The hypothetical temperature $T_{\rm hyp}(t_{\rm off})$ during the 
 sweep is then computed by matching 
\begin{equation}
E_\text{class}(t_{\rm off}) \equiv E_\text{class}^\mathrm{MC}(T_{\rm hyp}(t_{\rm off})) ,
\end{equation}
as illustrated in \figref{fig:SMTeff}.

\begin{figure}
	\includegraphics[width=\columnwidth]{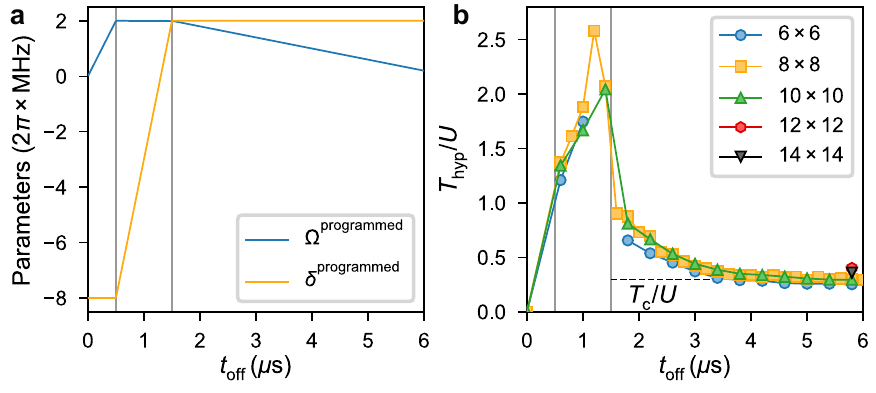}
	\caption{\textbf{Evolution of the hypothetical temperature.} \textbf{a,} Programmed state preparation protocol and \textbf{b}, corresponding hypothetical temperature $T_{\rm hyp}(t_{\rm off})$ during the sweep for different system sizes. The dashed line shows the classical critical temperature $T_{\rm c}$ for an infinite system with disorder averaged $U/\hbar = 2\pi \times 1.86\ \rm{MHz}$ for $\delta_{\rm f} = 2\pi \times 2 \ \rm{MHz}$.}
	\label{fig:SMExpT}
\end{figure}

In \figref{fig:SMExpT} we show $T_{\rm hyp}$ during the experimental state preparation process for different system sizes. At $t=0$
the system is in the ground state of $H_{\rm class}$ with $T_{\rm hyp}=0$, since we only consider arrays without any vacancies, and all atoms are in their ground state. During the first parts of the protocol where $\Omega$ is turned on and $\delta$ is increased to its final value, the hypothetical temperature strongly increases. 
This increase is mainly related to $H_{\rm class}$ not fully describing the quantum driving Hamiltonian $H_{\rm Ryd}$.

During the ramp-down process ($t_{\rm off} > 1.5 \ \mu \rm{s}$, $\delta = \delta_\text{f}$) the boundary to the AF phase is crossed, and we observe a strong ``cooling'' of the system. At the end of the sweep, we reach very low hypothetical temperatures close to or even below the infinite system size critical temperature $T_c$ (see following subsection). 
The finite temperature at the end of the sweep reflects that the state preparation is not perfectly adiabatic with the durations possible in the experiment.

\subsubsection{Classical critical temperature}
\begin{figure}
	\includegraphics[width=0.86\columnwidth]{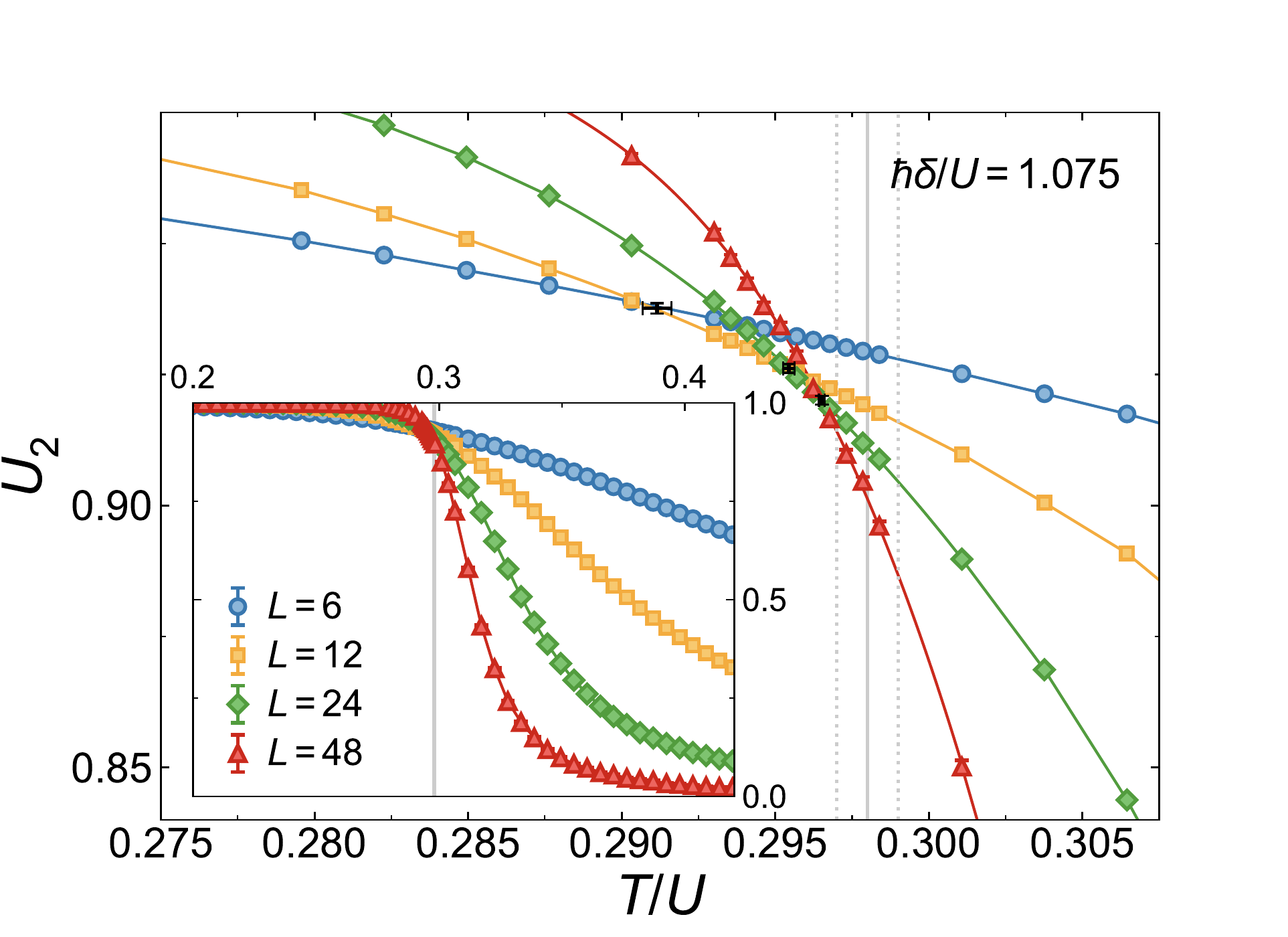}
        \caption{\textbf{Calculating the critical temperature.} The Binder cumulant $U_2$ and its ($L$, 2$L$) crossing points (black markers) for different linear system dimensions $L$, which allow to estimate the critical temperature in the thermodynamic limit. The solid (dotted) grey lines indicate the finite size extrapolated $T_c$ and its standard error.}
	\label{fig:SMbinder}
\end{figure}

To provide a natural scale for $T_{\rm hyp}$ we compute the critical temperature for the Hamiltonian parameters at the end of the considered sweep ($\Omega=0$), which is defined as the temperature below which the classical system is ordered in the thermodynamic limit. The critical temperature ${T_c/U = 0.298(1)}$ is extracted from the finite size extrapolation of the Binder cumulant ${U_2 = \frac{3}{2}\left(1-\langle m_\text{stag}^4\rangle / 3\langle m_\text{stag}^2\rangle^2\right)}$ crossing points as presented in Fig.~\ref{fig:SMbinder} \cite{Binder81}. Simulations to obtain the Binder cumulant have been performed for systems with homogeneous $\hbar\delta/U = 1.075$ (corresponding to $\delta_f=2\pi \times 2 \ \rm{MHz}$), again using 500 samples of random atom positions with an average nearest neighbour interaction $U/h = U_{\rm nnb}/h = 2\pi \times 1.86 \ \rm{MHz}$, a Euclidean interaction range of $5.2 a$ and periodic boundaries. 
Note that the scale $T_c$, which is only well defined in the infinite-size system, is provided as a reference, but does not preclude the emergence of long range order in finite size systems as observed.

\subsubsection{MPS time evolution versus classical equilibrium}\label{SM:QvsC}

\begin{figure}
	\includegraphics[width=8.5 cm]{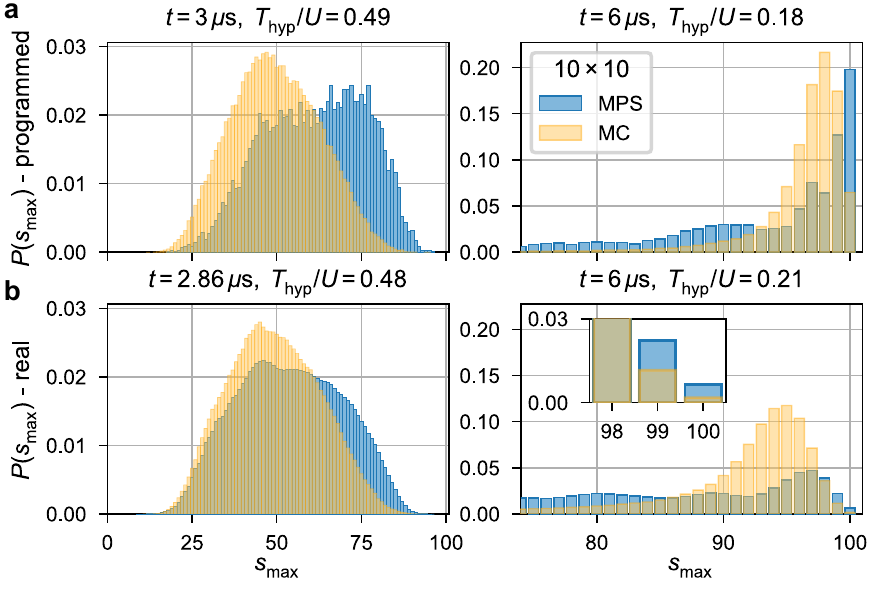}
	\caption{\textbf{Classical versus quantum state preparation.} Distribution of the largest N\'eel cluster sizes $s_{\rm max}$ during (left panel) and at the end (right panel) of the time evolution from MPS simulations (blue) compared to the classical equilibrium results (yellow) on a $10\times10$ square lattice. 
	\textbf{a}, Simulation in the setup without including any experimental imperfections. 
	\textbf{b}, Simulation including the known experimental imperfections.
	}
	\label{fig:SMmpsmc}
\end{figure}

\begin{figure}
	\includegraphics[width=8.5 cm]{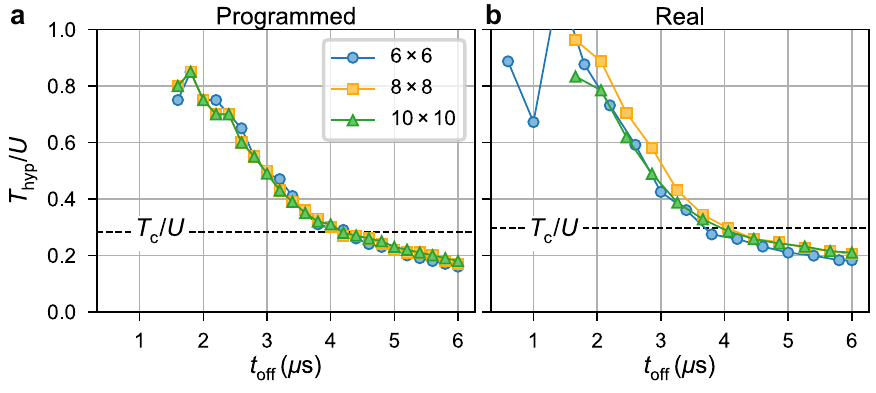}
	\caption{\textbf{Hypothetical temperature evolution for MPS.} $T_{\rm hyp}(t_{\rm off})$ for the MPS simulations during the final part of the state preparation protocol. Different symbols/colours show different system sizes. The panels show MPS simulations \textbf{a}, without, and  \textbf{b} with experimental imperfections.}
	\label{fig:SMmpsteff}
\end{figure}

We show in \figref{fig3} in the main text, that the distribution of AF domain sizes during the experimental time evolution is strikingly different from a classical equilibrium distribution with an assigned hypothetical temperature $T_{\rm hyp}$. Here, we support this result with an identical analysis for corresponding MPS simulations. We compare the distributions from MPS time evolutions to the corresponding classical equilibrium distributions from MC simulations for both the programmed setup, see \figref{fig:SMmpsmc}a, and the real setup, including experimental imperfections, see \figref{fig:SMmpsmc}b.
For intermediate times, shortly after entering the AFM dome (left panels), we observe a similar behaviour as in the experiment for both setups, where the distribution of the MPS simulations is generally shifted towards larger cluster sizes $s_{\rm max}$ compared to the classical distribution.
At the end of the sweep (right panels) we still observe that the simulations are not thermalised, since the MPS distributions are not matched by the classical equilibrium ones.
Also, the probability of creating N\'eel states is enhanced compared to the classical distribution, as was observed in the experiment. This effect is very strong in the programmed case, indicating that with experimental improvements we will gain in ground state preparation fidelity. 

In \figref{fig:SMmpsteff}, we show the evolution of $T_{\rm hyp}$ for the MPS simulations during the final part of the state preparation protocol for systems with up to $10 \times 10$ atoms for both protocols (programmed, real). Similar to the experiment, we observe a strong cooling of the system. At the end of the sweep, we observe that the extracted temperatures from the MPS simulations are somewhat lower than the ones from the experiment, for both the programmed and real cases, pointing to possible further imperfections not yet taken into account. 

To conclude one might wonder why the high probability of finding N\'eel states in the right panel of \figref{fig:SMmpsmc}a does not translate into a substantially lower hypothetical temperature $T_{\rm hyp}$ in
the left panel of \figref{fig:SMmpsteff}. The main reason is the temperature dependence of the specific heat of the equilibrium system. In the gapped AF phase the specific heat is exponentially suppressed at low temperature
$C(T)\sim \exp(-U/\mathrm{k_B}T)$~\cite{Onsager1944}. On the other hand the excess energy $\Delta E \equiv E_\text{class} - E_\text{class}^\mathrm{GS}$ is related to the integral of $C(T)$: 
$$\Delta E = \int_0^T C(T') dT'\ .$$ Combining the two formulas, we infer that $T\sim 1/\log (1/\Delta E)$, i.e. the matching temperature tends to zero only logarithmically slowly with $\Delta E$ going to zero.

\subsection{Triangular geometries}

In this section we consider the peculiarity of the finite-sized triangular array, leading to the observed reduction in the $m_{\rm stag}$ histogram observed for the $2/3$ phase. We also compare classical equilibrium calculations to experimental results on the triangular lattice.
\subsubsection{Triangular $2/3$ plateau}
\label{SM:tri2over3}

\begin{figure}
\includegraphics[width=8.5 cm]{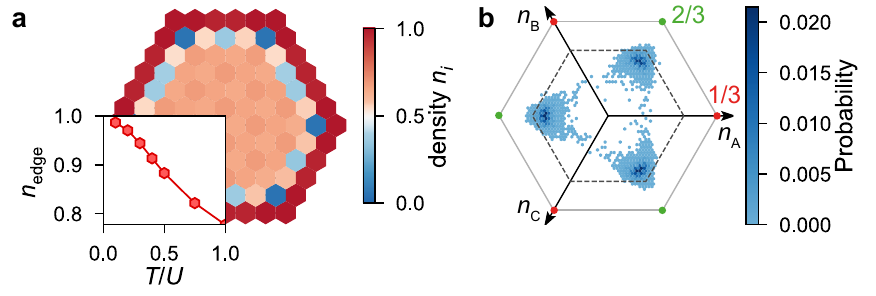}
\caption{\textbf{2/3 $m_{\rm stag}$ histogram for finite clusters.}
    MC results for a 108-site triangular cluster in the $2/3$ plateau 
with $\hbar\delta/U=4$ and temperature $T/U=0.1$. 
\textbf{a}, The real space Rydberg density $n_i$ shows that the outermost 
shell becomes fully populated at low temperature, as also illustrated 
in the inset which shows the Rydberg density at the edge. 
\textbf{b}, The corresponding sublattice magnetisation histogram does not 
reach its full potential width (outer hexagon), since the edge sites cannot 
participate in the formation of the $2/3$-filling states. 
The dashed hexagon shows the maximal extend for the histogram 
when only the sites of the system without the edge are considered.}
\label{fig:SM2over3}
\end{figure}
\begin{figure}
	\includegraphics[width=8.5 cm]{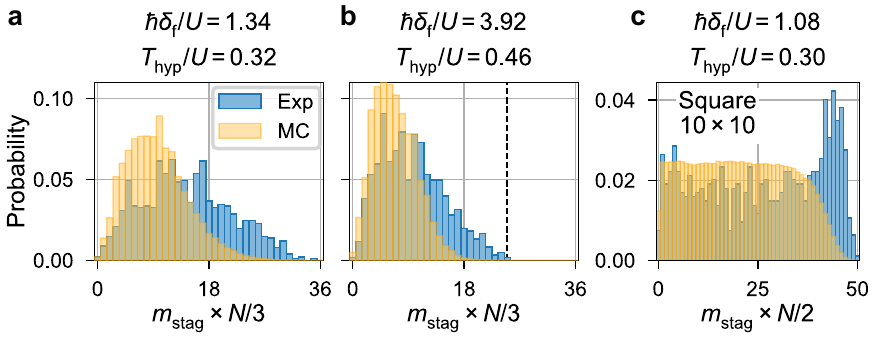}
	\caption{\textbf{Quantum real-time evolution versus classical equilibrium on the triangular lattice.} 
		\textbf{a, b}, The distribution of the triangular order parameter $m_{\rm stag}$ at the end of the state preparation protocols entering the $1/3$ (\textbf{a}) and $2/3$ (\textbf{b}) regimes.
		\textbf{c}, Distribution of $m_{\rm stag}$ for a $10 \times 10$ square lattice at the end of the sweep entering the AF phase, as a comparison. 
		In all panels, the blue (yellow) bars show experimental (corresponding classical) results. The dashed line in \textbf{b} shows the maximal value of $m_{\rm stag}$ in the $2/3$ regime induced by the cluster boundaries.}
	\label{fig:SMTriRadHists}
\end{figure}

The edge of the finite-sized clusters can have an impact on the achievable ground states, which might differ from the ones expected from the corresponding bulk phase diagram. 
In particular, the edge sites have fewer direct neighbours than the bulk sites and, hence, 
they can be excited to Rydberg states at lower $\delta/U$.
For the results in this paper, this effect is particularly important for the $2/3$ phase on the triangular clusters, as we will now discuss.

In \figref{fig:SM2over3} we show MC results for a fixed $\hbar\delta/U=4$ in 
the $2/3$-filling regime at low temperature $T/U=0.1$. 
It is clearly visible in the Rydberg density $n_i$, that the edge sites 
become completely filled with Rydberg states in the ground state 
$T\rightarrow0$. In particular, the Rydberg density on the edge 
$n_{\rm edge}$ approaches one when the temperature is reduced, 
see inset in \figref{fig:SM2over3}a. 
Therefore, only the atoms in the bulk can take part in the formation of 
the $2/3$-filled states, and the maximal order parameter is reduced. 
This can be directly observed in the sublattice magnetisation histogram, 
see \figref{fig:SM2over3}b, which shows that the position of the 
three peaks defining the $2/3$ phase do not reach their maximal 
possible value at the edge of the outer hexagon. Instead, 
they are bounded by the dashed hexagon, which shows the 
maximal extend of the histogram for the bulk system.
The reduction of the maximal radius is rather large, since it scales with the number of sites on the boundary. In particular, for the here shown 108-site cluster, the boundary consists of 33 sites, leading to $\sim30\ \%$ reduction of the maximal radius in the histogram. This effect isn't observed in the $1/3$ regime as $\delta_f$ is small enough that the boundary sites also obey the AF ordering. \\

\subsubsection{Time evolution vs classical equilibrium on the triangular lattice}
\label{SM:MCvsExpTri}

Here, we demonstrate that on the triangular lattice the experimental state preparation protocol, again, cannot be reproduced by a classical equilibrium distribution. 
In \figref{fig:SMTriRadHists}a, b we show the experimentally obtained distribution of the order parameter, $m_{\rm stag}$, at the end of the sweep preparing the (a) $1/3$ and (b) $2/3$ phases for the 108~atom array, and compare them to the corresponding classical distribution with hypothetical classical temperature $T_{\rm hyp}$, assigned as described in Sec.~\ref{SM:T_hyp}.
The experimental distribution is centred at substantially larger values of $m_{\rm stag}$, demonstrating an increased probability of finding highly ordered states compared to a corresponding classical annealing.
In \figref{fig:SMTriRadHists}c we show, as a comparison, the distribution of $m_{\rm stag}$ for a $10 \times 10$ square lattice at the end of the sweep preparing the AF phase, 
which shows similar, but enhanced, features.

\end{document}